\documentclass[aip,reprint]{revtex4-1}

\usepackage[utf8]{inputenc} 
\usepackage[T1]{fontenc}    
\usepackage{hyperref}       
\usepackage{url}            
\usepackage{booktabs}       
\usepackage{amsfonts}       
\usepackage{nicefrac}       
\usepackage{microtype}      
\usepackage{graphicx}
\usepackage{lipsum}
\usepackage{comment}
\usepackage[caption=false]{subfig}
\usepackage{textcomp}
\usepackage{soul}           
\usepackage{amsmath}
\usepackage{listings}
\usepackage{chngcntr}
\usepackage{enumitem}
\usepackage[table,xcdraw]{xcolor}
\usepackage{booktabs}
\usepackage{multirow}
\usepackage{hhline}

\usepackage{euler}

\newcommand{\enrico}[1]{{#1}}

\definecolor{codegreen}{rgb}{0,0.6,0}
\definecolor{codegray}{rgb}{0.5,0.5,0.5}
\definecolor{codepurple}{rgb}{0.58,0,0.82}
\definecolor{backcolour}{rgb}{0.95,0.95,0.92}

\lstdefinestyle{mystyle}{
    backgroundcolor=\color{backcolour},   
    commentstyle=\color{codegreen},
    keywordstyle=\color{magenta},
    numberstyle=\tiny\color{codegray},
    stringstyle=\color{codepurple},
    basicstyle=\ttfamily\footnotesize,
    breakatwhitespace=false,         
    breaklines=true,                 
    captionpos=b,                    
    keepspaces=true,                 
    numbers=left,                    
    numbersep=5pt,                  
    showspaces=false,                
    showstringspaces=false,
    showtabs=false,                  
    tabsize=2
}
\let\oldtextbf=\textbf
\renewcommand*{\textbf}[1]{\ifmmode\mathbf{#1}\else\oldtextbf{#1}\fi}
\renewcommand*{\phi}[0]{\varphi}

\draft 

\begin{document}

\title[]{Computing the Committor with the Committor: an Anatomy of the Transition State Ensemble} 



\author{Peilin Kang}
\affiliation{Atomistic Simulations, Italian Institute of Technology, 16156 Genova, Italy}

\author{Enrico Trizio}
\affiliation{Atomistic Simulations, Italian Institute of Technology, 16156 Genova, Italy}
\affiliation{Department of Materials Science, Università di Milano-Bicocca, 20126 Milano Italy}

\author{Michele Parrinello}
\email[]{michele.parrinello@iit.it}

\affiliation{Atomistic Simulations, Italian Institute of Technology, 16156 Genova, Italy}


\date{\today}

\begin{abstract}
The study of the kinetic bottlenecks that hinder the rare transitions between long-lived metastable states is a major challenge in atomistic simulations. We propose a method to explore the transition state ensemble, which is the distribution of configurations that the system passes as it translocates from one metastable basin to another. We base our method on the committor function and the variational principle to which it obeys. We find its minimum through a self-consistent procedure that starts from information limited to the initial and final states. Right from the start, our procedure allows sampling very many transition state configurations. With the help of the variational principle, we perform a detailed analysis of the transition state ensemble, ranking quantitatively the degrees of freedom mostly involved in the transition and opening the way for a systematic approach for the interpretation of simulation results and the construction of efficient physics-informed collective variables. 
\end{abstract}


\maketitle 

\section{Introduction}
    Many important physicochemical transformations like crystallization, chemical reactions, and protein folding take place on a time scale that is not directly accessible to microscopic simulations.  
    These processes are referred to as rare events and are hindered by kinetic bottlenecks that slow the rate of transition between metastable states. 
    Such kinetic bottlenecks are present whenever the metastable states are separated by a high free-energy set of configurations that we refer to as the transition state ensemble (TSE).
    Finding and analyzing this region is of the utmost theoretical and practical importance.
    For example, identifying the transition state is considered the holy grail when it comes to chemical reactions~\cite{solomons2008organic}, as it provides precious information about reaction mechanisms and rates~\cite{anslyn2006modern}, or when dealing with proteins, as it can provide \enrico{insight into} their dynamics~\cite{baldwin1999protein, cecconi2005, jackson1998}.

    In the vast rare event literature~\cite{torrie1977nonphysical,laio2002escaping,izrailev1999steered,darve2001calculating,sugita1999replica,Henin2022enhanced}, the determination of the TSE is usually the culmination of the simulation. 
    Here, instead, we make the determination of the transition state ensemble the first and key aspect of our investigation.  
    The theoretical tool that allows this change of perspective is the committor function  $q(\textbf{x})$, introduced by Kolmogorov~\cite{kolmogoroff1931analytischen}. 
    Given two metastable states, A and B, $q(\textbf{x})$ gives the probability that starting from configuration  $\textbf{x}$, the system ends in B without having first passed by A \enrico{and, as a consequence, it is conventionally used to identify the TSE as the set of configurations for which $q(\textbf{x})\simeq\frac{1}{2}$.}~\cite{onsager1938initial,berezhkovskii2005one,berezhkovskii2013diffusion,du1998transition,weinan2005transition,Dellago2006,vanden2006transition,weinan2010transition,roux2022transition,jung2023machine,rotskoff2020learning,chen2023discovering}  
    Once $q(\textbf{x})$ is known, properties like the transition rate between A and B, the density of reactive trajectories, or the reactive fluxes can be computed.~\cite{vanden2006transition,weinan2005transition,weinan2010transition} 
    \enrico{Unfortunately, the determination of $q(\textbf{x})$ is challenging, and, in the transition path sampling literature,~\cite{bolhuis2002throwing, Dellago2006,jung2023machine} it has been mostly estimated for curated sets of points via the \emph{committor analysis}. However, such an approach can be computationally expensive and often dependent on the choice of the criteria used to determine whether a given trajectory is committed to either basin A or B~\cite{lazzeri2023committor}}.
    
    \enrico{An alternative to such an approach is based on the theory of Kolmogorov.~\cite{weinan2010transition,khoo2019solving,li2019computing,chen2023committor}. who} has shown that $q(\textbf{x})$ can be determined as the solution of a partial differential equation~\cite{weinan2010transition} that obeys the boundary conditions $q(\textbf{x}_A)=0$ and $q(\textbf{x}_B)=1$ where $\textbf{x}_A$ and $\textbf{x}_B$ denote two configurations belonging to basin $A$ and $B$, respectively. Unfortunately, solving such a multidimensional equation for real systems poses insurmountable problems. 
    However, \enrico{ under the hypothesis of overdamped dynamics,} the solution of the Kolmogorov equation can also  be obtained following a variational approach that amounts to minimizing a functional of the committor $\mathcal{K}[q(\textbf{x})]$ which, neglecting immaterial multiplicative constants, can be written as: 
        \begin{equation}
            \mathcal{K}[q(\textbf{x})] =\Big \langle \big|\nabla q(\textbf{x})\big|^2 \Big\rangle_{U(\textbf{x})} 
            \label{eq:variational_functional}
        \end{equation}
    where in the differential operator, the derivatives pertinent to the $i^{th}$ atom of mass $m_i$ are performed with respect to its mass-weighted Cartesian coordinates $\sqrt{m_i}\textbf{x}_i$, the average is over the Boltzmann ensemble driven by the interaction potential $U(\textbf{x})$ at the inverse temperature $\beta$, and the boundary conditions $q(\textbf{x}_A)=0$ and $q(\textbf{x}_B)=1$ are implied. Moreover, the reaction rate $\nu_R$ is proportional to the minimum of the functional  $\mathcal {K}_m$. \enrico{For further discussion on Eq.~\ref{eq:variational_functional} and its extension to the general Langevin dynamics, we refer the Reader to the SI, Sec.~\ref{sup_sec:variational_principle}, and the excellent review by Weinan E and Eric Vanden-Eijnden.~\cite{weinan2010transition}}  
 
    Unfortunately, even when using the variational approach, evaluating $q(\textbf{x})$ remains challenging.
    In order to understand this sampling difficulty, we notice that \enrico{when dealing with rare events}, trajectories started in A have a very small probability of ending in B, thus $q(\textbf{x})\approx 0$ for $\textbf{x}\in A$, and similarly, when a trajectory is started in $B$ it will most likely remain in B, thus $q(\textbf{x})\approx 1$ for $\textbf{x} \in B$.  
    As a consequence, $|\nabla q(\textbf{x})|^2$ is significantly different from zero only in those regions in which $q(\textbf{x})$ goes from 0 to 1 as it passes through the transition state region, where is at its largest.
    This is hardly surprising since the probability of going from A to B, which $q(\textbf{x})$ reflects, is crucially determined by the TSE, the very set of configurations that are difficult to sample in a rare event scenario. 
    
    This difficulty has been recognized, and different remedies have been proposed. In Ref.~\cite{khoo2019solving}, it is shown that a uniform sampling can be done in small systems, but it is not, in general, a viable strategy. For this reason, a variety of enhanced sampling methods have been suggested to collect configurations that pertain to the TSE so as to estimate accurately $\mathcal{K}[q(\textbf{x})]$ and eventually minimize it. The methods used range from metadynamics~\cite{li2019computing} to a combination of umbrella sampling and parallel tempering.~\cite{rotskoff2020learning} If the objective is to compute $q(\textbf{x})$, this approach is wasteful since the TSE is a small portion of configuration space, and even \enrico{if one uses} enhanced sampling methods, it is rarely visited. Furthermore, the calculation of the committor comes at the end of what amounts to having solved the rare event problem, and since these methods are dependent on the choice of the collective variable, the accuracy of the result is, at times, difficult to assess.
 
    Methods that are similar in spirit to ours are described in \enrico{Ref.~\cite{krivov2021nonequilibrium}, where short nonequilibrium trajectories are harvested iteratively closer to the TS, and in }Refs.~\cite{chen2023discovering,jung2023machine,lazzeri2023committor}, which are based on path sampling and thus, like ours, are focused from the start on sampling the region close to the TSE. However, these methods are somewhat complex and can only be applied if the path sampling approach is at least in part successful. Thus, similarly to the other approaches described above, all these methods require that the rare event problem is at least partially already solved\enrico{, a limitation also shared by the method presented in Ref.~\cite{krivov2018committor}, which allows computing the $q(\textbf{x})$ from long equilibrium trajectories}. 
     
\begin{figure*}[t!]
        \centering
        \includegraphics[width=\linewidth]{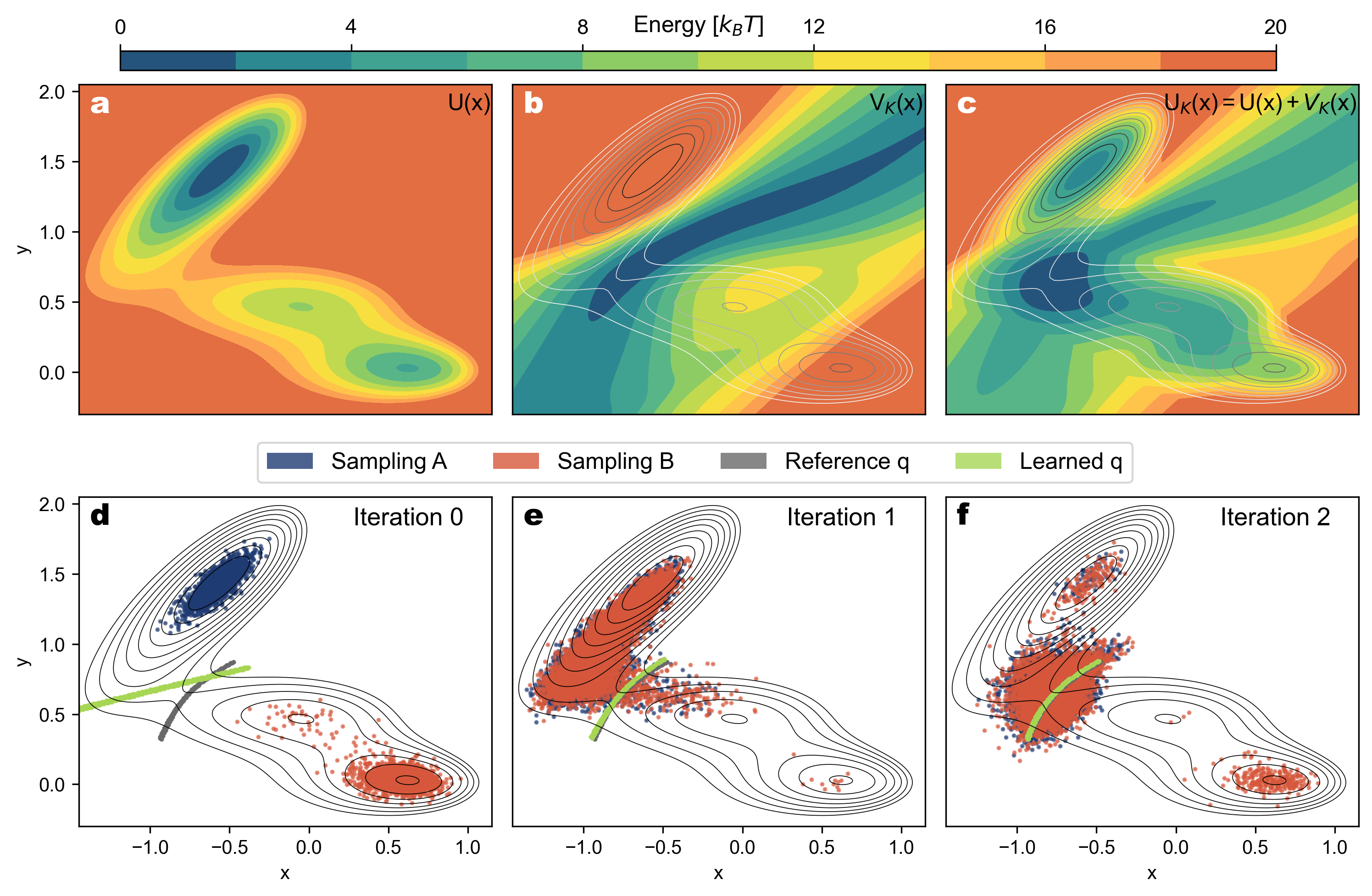}
        \caption{We illustrate how our iterative procedure proceeds when applied to the computation of the committor function of the  M\"uller-Brown potential.  
        In \textbf{a}, \textbf{b} and \textbf{c} we report the contour plots of the potential $U(\textbf{x})$, the bias $V_\mathcal{K}(\textbf{x})$  and the effective biased potential $U_\mathcal{K}(\textbf{x}) = V_\mathcal{K}(\textbf{x})+U(\textbf{x})$, respectively.
        Here, the committor is computed from numerical integration. 
        The scatter plots in panels \textbf{d},\textbf{e},\textbf{f} report the results of the sampling cycles through the iterations needed to converge. Points sampled starting from basin A are depicted in blue, whereas those from B are in red. 
        For each iteration, the \enrico{$\simeq$0.5 value of} the learned committor function (green line) is compared with the reference value from numerical integration (grey line) for the physically relevant part of the transition state region. In the lower panels, the original potential energy surface is represented by isolines.}
        \label{fig:muller}
    \end{figure*}

    Instead, our approach relies only on the knowledge of the initial and final state and can be initiated just by performing unbiased simulations in the initial and final metastable basins. To sample the TSE, we use to our advantage what appears to be a handicap. To this effect, we introduce the following committor-dependent bias potential
        \begin{equation}
            V_\mathcal {K}(\textbf{x})=-\frac{1}{\beta}\log (|\nabla q(\textbf{x})|^2)
        \end{equation}
    It follows from the general behavior of $q (\textbf x)$ that such a bias is repulsive in $A$ and $B$, where $\nabla q (\textbf x) \approx 0$, and becomes highly attractive close to the TSE, where $q (\textbf x)$ raises very rapidly from $0$ to $1$. \enrico{In particular, the maximum value is reached for $q(\textbf{x})\simeq\frac{1}{2}$ (see Fig.~\ref{fig:muller} \textbf{b} and Fig.~\ref{sup_fig:visual_bias}), and, as a consequence, such a bias has the appealing property of driving the sampling towards the region which is conventionally associated with the TSE.
    In addition, we notice that the standard TSE definition can be enriched by explicitly taking into account also the probability that the configuration $\textbf{x}$ is actually visited, which is only implicitly considered in the standard approach. 
    Indeed, in that case, the points selected for computing the committor are generated in a transition path sampling run or even along actual reactive trajectories.}
    
    \enrico{Such considerations motivate us to define the TSE ensemble by what we call the \emph{Kolmogorov distribution}:
        \begin{equation}
            p_\mathcal{K}(\textbf x) = 
            \frac{e^{-\beta U_\mathcal{K}(\textbf{x}) }}{Z_\mathcal{K}} \quad\text{with}\quad U_\mathcal{K}(\textbf{x})=U(\textbf{x}) + V_\mathcal{K}(\textbf{x})
            \label{eq:kolmogorov_distribution}
        \end{equation}
    where $U_\mathcal{K}$ is the biased potential and $Z_\mathcal{K}=\int d\textbf{x} e^{-\beta U_\mathcal{K} (\textbf{x})}$ the corresponding partition function.}
    \enrico{Somehow reassuringly, $p_\mathcal{K}(\textbf{x})$ is also closely related to the Kolmogorov functional since $\mathcal{K}[q(\textbf{x})] = \frac{Z_\mathcal{K}}{Z_B}$, thus allowing rewriting it as:
        \begin{equation}
            \mathcal{K}[q(\textbf{x})]={\Big \langle \frac {1}{|\nabla q(\textbf{x})|^2}\Big\rangle ^{-1} _{U_{\mathcal K}}}  
        \end{equation}
    While equivalent to Eq.~\ref{eq:variational_functional}, this expression suggests that it would be more profitable to use $U_\mathcal K$ to generate the points needed to estimate statistically  $\mathcal{K}[q(\textbf{x})]$, since this approach automatically enhances a physically meaningful TSE sampling (see Fig.~\ref{fig:muller} \textbf{c}).}
    
    However, at first sight, this may still appear to be a chicken and egg problem since to get the committor, one needs good sampling, and in turn, to get good sampling, one needs the committor. We show here that this dilemma can be resolved by setting up the self-consistent iterative procedure, described in the method section. This procedure starts from an initial estimate of the committor. Such an estimate needs to have the property of being $ \approx 0$ for $\textbf{x} \in A$  and  $ \approx 1$ for $\textbf{x} \in B$ and of interpolating smoothly between the two basins. 
    One simple way of obtaining an initial guess is to express it as a classifier trained using data obtained by performing unbiased simulations in the two basins.  
  
    After convergence is reached, a large number of TSE configurations can be harvested, and the property of the TSE analyzed in great and illuminating detail. 
    This analysis is facilitated by the fact that we express $q(\textbf{x})$ as a neural network~\cite{ma2004automatic,khoo2019solving,li2019computing,chen2023committor} (NN) $q_\theta (d(\textbf{x}))$ whose weights are denoted by $\theta$ and whose input features $d(\textbf{x})$ are a set of physical descriptors that simplify the imposition of the problem symmetries, decrease the noise, and help understanding the physics of the problem.
    In particular, we shall use the approach from Ref.~\cite{bonati2020data} to rank the relevance of the descriptors.

    The variational principle also provides a powerful way of choosing the descriptors. In fact, the inclusion of physically relevant descriptors lowers the variational bound, while adding physically irrelevant descriptors has very little effect. 
    Moreover, in complex systems and at finite temperatures the TSE is not associated with only one structure and its quasi-harmonic excited vibrational states. Rather, it is populated by different competing structures.     
    To sort them out, we use the k-medoids clustering method~\cite{Schubert2022kmedoids},  which also associates to each cluster a \emph{medoid} configuration that best represents the cluster structure.
    The combination of these tools greatly facilitates the physical interpretation of the results and guides the researcher's attention to the degrees of freedom that matter the most. 
    It is to reflect the ability of our method to analyze the TSE in excruciating detail that we have chosen the manuscript title.
        
    We first test our method on the numerically solvable example of the Müller potential, and proceed then to discuss the classical example of alanine dipeptide in vacuum, a complex chemical reaction, and the folding of a small protein. In all the last three examples, our analysis leads to novel insight, even on a much-studied problem like that of alanine dipeptide.    
    The details of the numerical implementation are discussed in the method section.
    
\section {Results}
    \subsection{Müller-Brown Potential}
        The first application of our methods is to the two-dimensional Müller-Brown potential, which is often used to test new methods since, in this case, the committor can be numerically evaluated (see SI, Sec.~\ref{sup_sec:numerical_muller}).

    \begin{figure*}[t!]
        \centering
        \begin{minipage}{0.72\linewidth}
           \includegraphics[width=\linewidth]{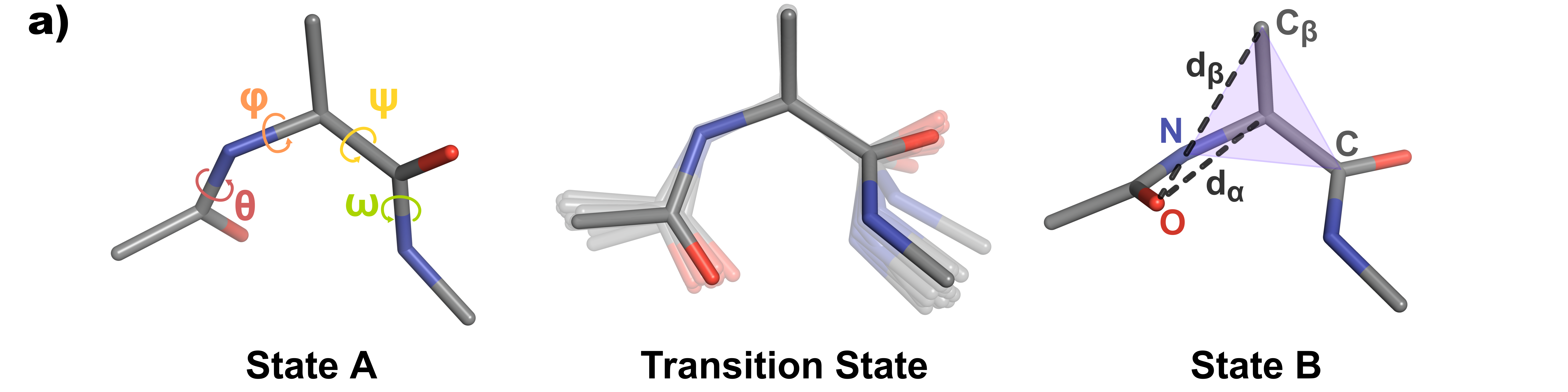}
           \hfill
            \includegraphics[width=\linewidth]{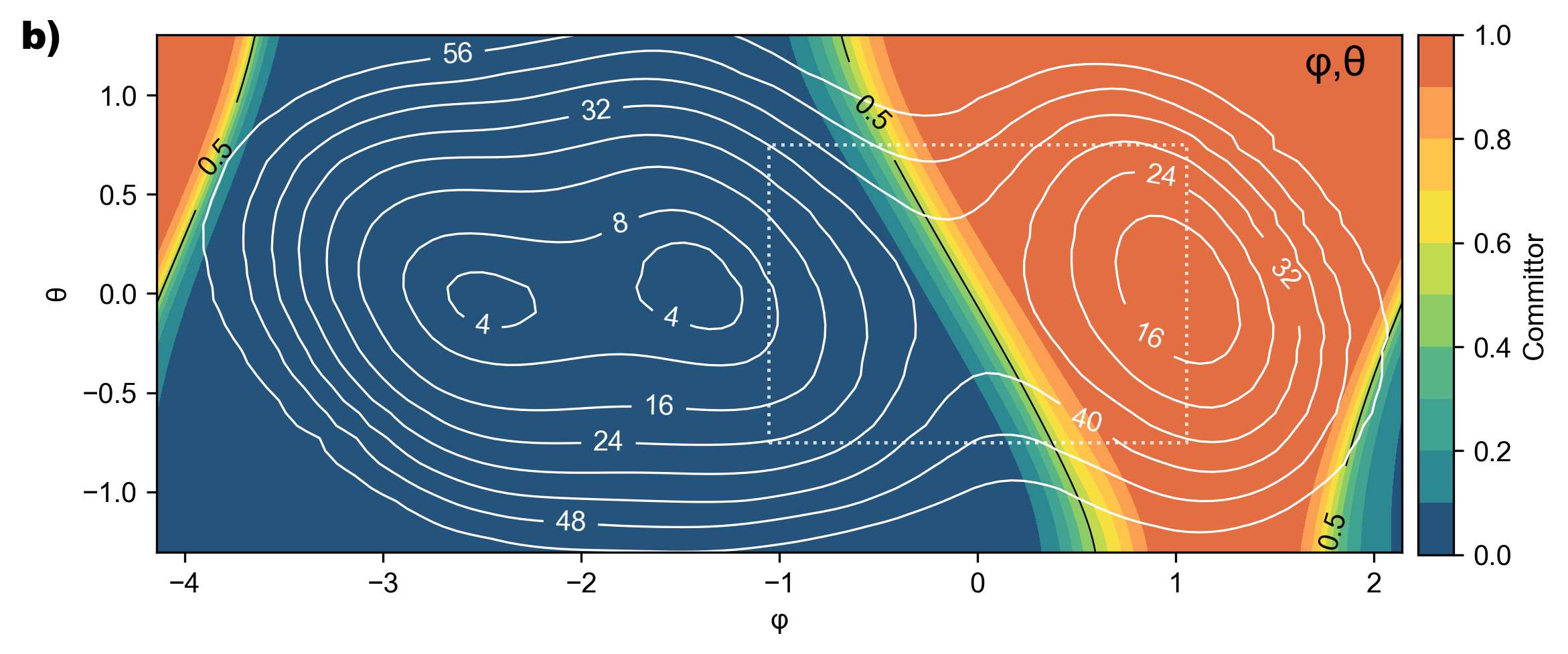} 
        \end{minipage}
        \begin{minipage}{0.26\linewidth}
           \includegraphics[width=\linewidth]{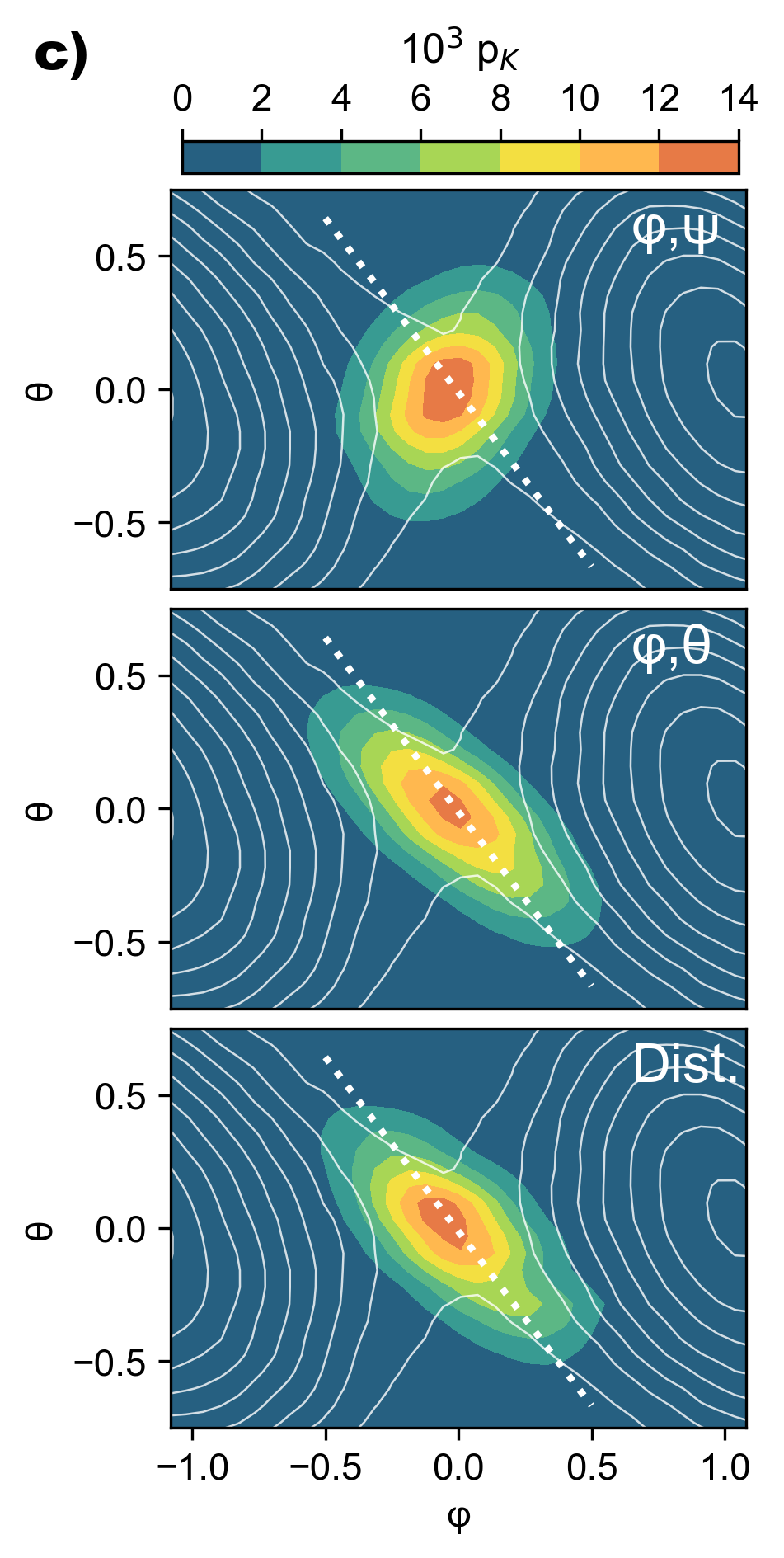} 
        \end{minipage}
        \caption{\textbf{a)} Snapshots of alanine dipeptide conformations in metastable and transition states. The relevant torsional angles of the peptide are depicted for state A, whereas the most relevant distances and the plane defined by the N,C$_\beta$,C atoms are illustrated for state B. For the transition state, we report in transparency a superimposition of 20 configurations from the TSE and, in solid color, the medoid of such an ensemble.
        \textbf{b)} Projection of alanine FES (white isolines) and of the learned committor value $q(\phi, \theta)$ (colormap) in the $\phi,\theta$ plane.
        \textbf{c)} Contour plot the Kolmogorov distribution $p_\mathcal{K}$ on the transition region, highlighted in \textbf{b} by a dotted white box, according to committor-based bias potentials trained on different descriptors sets $\phi,\psi$ angles, $\phi,\theta$ angles and a set of distances, as indicated by the top-right labels. The white isolines depict the underlying FES, and the white dashed line reports the linear relation between $\phi,\theta$ for the TSE configurations proposed in Ref.\cite{bolhuis2000reaction}.}
        \label{fig:alanine}
    \end{figure*}

        In the bottom panels of Fig.~\ref{fig:muller}, we show the evolution of the segment of the isocommittor line $q(\textbf{x})=\frac{1}{2}$ that overlaps the TSE and compare it to the result coming from numerical integration.   
        In the first iteration, in which the only data available are those coming from the metastable states, the first guess for $q(\textbf{x})$ is nothing but a classifier. 
        Thus, at this first stage, the isocommittor line is just a straight line that divides the two metastable basins. 
        However, as more and more TSE data are collected under the action of the bias $V_\mathcal{K}(\textbf{x})$, the isocommittor converges to its correct value after only a few iterations, and an accurate description of the whole $p_\mathcal{K}(\textbf{x})$ is obtained after only a few more iterations (see SI, Fig.~\ref{sup_fig:muller_kolmogorov_comparison}).
        Moreover, it can be seen that as the iterative process progresses, the TSE is better and better sampled.

    \subsection{Alanine Dipeptide}
        As a second and more physically relevant example, we study the transition in vacuum of alanine dipeptide between the $C_{7eq}$ (A) and  $C_{7ax}$ (B) conformers, which is one of the most studied rare event models.
        Its conformational landscape is spanned by the four dihedral angles $\phi$, $\theta$, $\psi$, and $\omega$ that measure the orientation of the two peptides relative to the more rigid tetrahedron formed by the N, C, C$_{\alpha}$, and C$_{\beta}$ atoms (see panel \textbf{a} of Fig. \ref{fig:alanine}).
        The dihedral angles $\phi$ and $\psi$ have been found to be good collective variables when used in enhanced sampling methods.~\cite{bonati2020data}  
        It has also been found in transition path sampling studies~\cite{bolhuis2000reaction} that in the configurations that belong to the TSE, the angles $\phi$ and $\theta$ are, modulus a constant, approximately linearly anti-correlated (i.e., $\theta \simeq - \,\phi$). 
        In order to demonstrate the ability of our method to recover these results, we first use $\phi,\theta, \psi,\text{and}\,\, \omega $ as descriptors.
        But rather than using all of them at once, we start with one, and then we systematically add all the others.
     
        The rationale for this procedure is that, given the variational property of the functional $\mathcal{K}[q(\textbf{x})]$, an indication of the relevance of an added dihedral angle will be its ability to lower the minimum value $\mathcal{K}_m$.
     
        The number of calculations to be performed is reduced if we first note that, in order to satisfy the boundary conditions, $\phi$ has to be part of the descriptor set.
        Thus, we compute $\mathcal{K}_m$ first using only $\phi$ and then study all possible combinations of $\phi$ with the remaining torsional angles.
        The results of these calculations are illustrated in Table~\ref{sup_tab:alanine_descriptors} in the SI, where it can be seen that including $\theta$ in the descriptors set is by far the most effective in lowering $\mathcal{K}_m$, confirming that $\theta$ is a crucial part of the reaction coordinate~\cite{bolhuis2000reaction}.

    \begin{figure*}[t!]    \centering\includegraphics[width=0.8\linewidth]{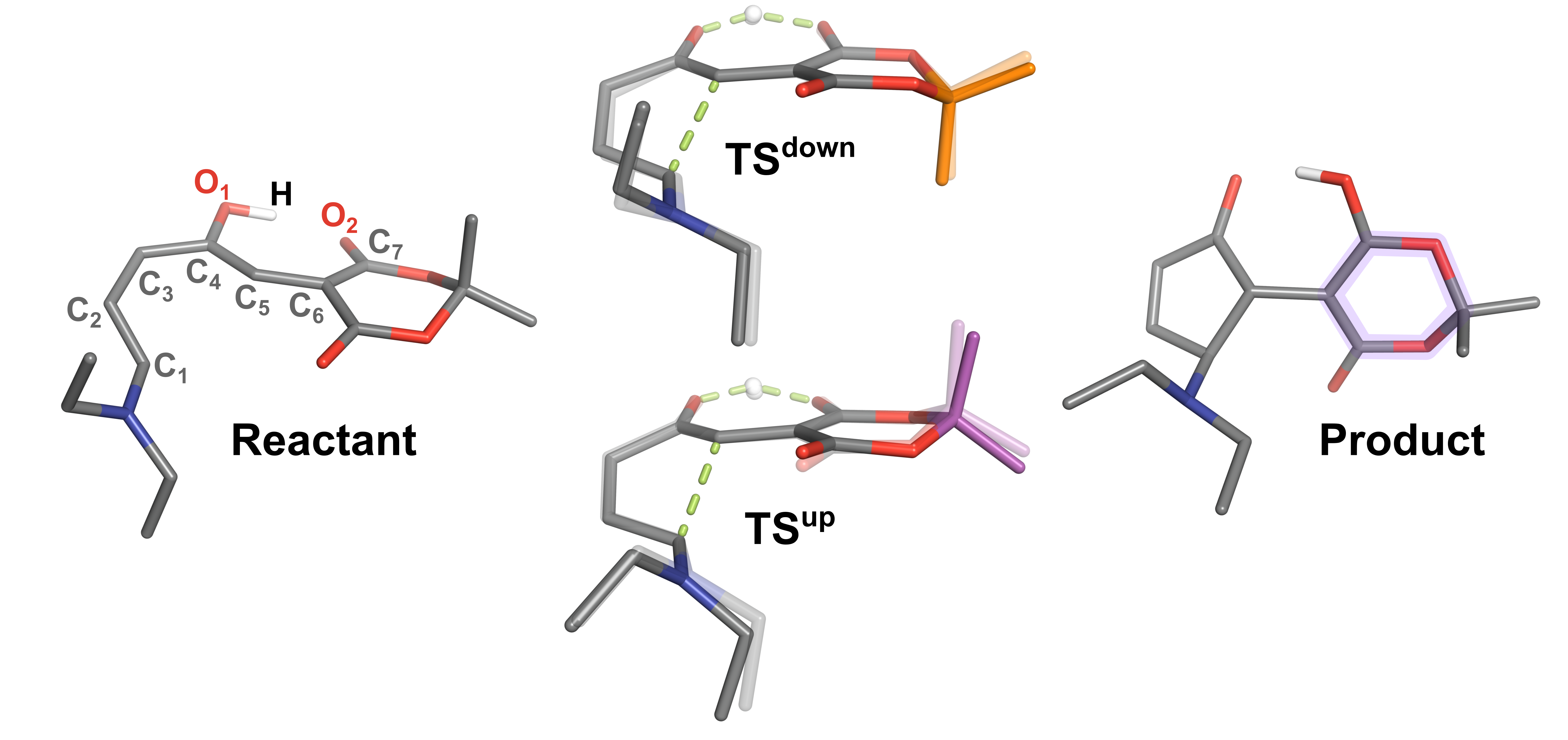}
        \caption{Snapshots of reactant, product and transition species involved in DASA reaction.
        On the reactant configuration, the labels highlight the atoms used to compute the distance-based descriptors set used for the training of the committor model, whereas the purple shadow on the product highlights the 1,3-dioxane ring used to compute the puckering coordinates.
        For the transition state (TS), we report two configurations that are representative of the TSE according to clustering and puckering analysis. In solid color, we report the medoids of each cluster, and in transparency, the corresponding reference configuration obtained via dimer method\cite{henkelman1999dimer}. The atoms that characterize the TS$^{down}$ configuration are highlighted in orange, whereas the TS$^{up}$ ones are in purple.
        For such configurations, the bonds that change through the reaction are shown as dashed green lines.}
        \label{fig:dasa}
    \end{figure*}

        Furthermore, the approximate linear relation between $\theta$ and $\phi$ in the TSE reported in Ref.~\cite{bolhuis2000reaction} is observed if and only if $\theta$ is included in the descriptors set. 
        This point is illustrated in two top plots in panel \textbf{c} of Fig.~\ref{fig:alanine}, where the marginal of the $p_\mathcal{K}$ TSE distribution relative to $\phi$ and $\theta$ is drawn on the $\phi$ and $\theta$ free energy surface (see also SI, Fig.~\ref{sup_fig:alanine_TS_from_q}).
        In all cases, in this representation, the marginal has an elongated ellipsoidal structure whose main axis is misaligned relative to the expected behavior if $\theta$ is not part, explicitly or implicitly, of the descriptor set.  
        This shows that the combination $\phi, \psi$, although efficient when used in an enhanced sampling context, does not fully capture the nature of the TSE. 
        As a corollary to this analysis, we agree with Ref.~\cite{bolhuis2000reaction} that the alanine dipeptide free energy surface should be more expressively represented if drawn as a function of $\phi$  and $\theta$ as in Fig.~\ref{fig:alanine} rather than using the standard $\phi,\psi$ projection since it brings out clearly the role of $\theta$.

        \enrico{This instructive example also allows us to compare in the practice our definition of the TSE based on $p_\mathcal{K}$ with the conventional one based on the $q(\textbf{x})\simeq\frac{1}{2}$. 
        Indeed, the latter also includes unlikely configurations with extremely high energy, whereas our criterion only focuses on the physically relevant region (see also SI, Fig.~\ref{sup_fig:alanine_pk_vs_q05}).}
    
        In the alanine dipeptide case, we had enough prior knowledge of the system that we could solve the problem using a reduced set of descriptors.
        However, when one approaches a new system, this is rarely the case.
        For this reason, as a demonstration of the possibilities of our method, we also take a blind approach and assume that we only know the initial and final conformations. 
        For this physics-agnostic calculation, we use as descriptors the 45 distances between the alanine heavy atoms. 
        At convergence, we find that this descriptor set does much better than the ones based on dihedrals only, reaching the value of $\mathcal{K}_m= 1.1$ a.u., where a.u. stands for arbitrary units, and the TS linear $\theta$-$\phi$ correlation is respected, as shown in the bottom of panel \textbf{c} of Fig.~\ref{fig:alanine}, since the $\theta$ degree of freedom is taken into account, albeit implicitly.
        This is not surprising, given the much higher variational flexibility of the trial committor function. 
        
        However, in so doing, we lose physical transparency, and the price for this unbiased generality is that further analysis is needed~\cite{novelli2022lasso}. To this effect, we use a tool exploited in Ref.~\cite{bonati2020data} that allows ranking the descriptors according to their weight in the optimized $q_\theta(\textbf{x})$ model (see SI, Sec.~\ref{sup_sec:feature_ranking}). 
        From this ranking (see SI, Fig.~\ref{sup_fig:alanine_rank}), it emerges that the two distances  $d_{\alpha}$ and $d_{\beta}$ (see panel \textbf{a} of Fig.~\ref{fig:alanine}) stand out as the most relevant. This might seem at first surprising, but we note these two distances reflect the position of O relative to the plane that passes through  N, C, and  C$_{\beta}$.    
        In turn, the position of O depends on $\theta$ and $\phi$, thus the prominence of $d_{\alpha}$ and $d_{\beta}$ is a way in which the NN expresses the TSE conformation using only interatomic distances.  
        Of course, the dihedrals are the natural language in which to describe a conformational change, while the description in terms of interatomic distances is less immediately evident, but it is reassuring that the physical conclusions are the same even if cast in a different language.

    \begin{figure*}[t!]
        \centering
        \includegraphics[width=\linewidth]{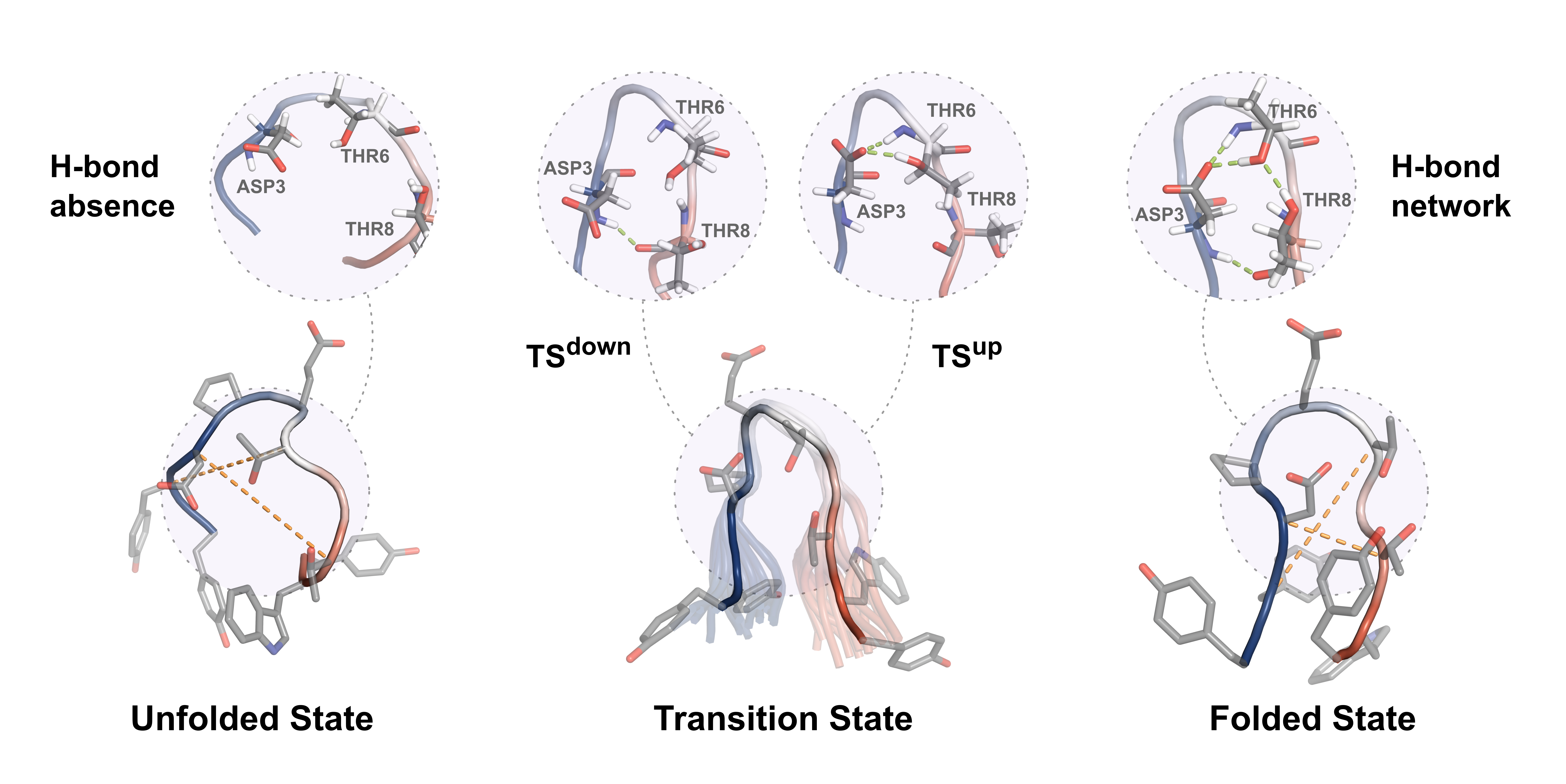}
        \caption{Snapshots of representative configurations from the folded (F), unfolded (U), and transition (TS) states (bottom row) of chignolin protein and the corresponding hydrogen bonds (top row). In the bottom row, the main chain of the protein is depicted as a cartoon colored from blue to red according to the residue number, whereas the side chains are reported in transparency in licorice. 
        For the F and U metastable states, we report as orange dashed lines the distances C$_\alpha^2$C$_\alpha^6$ and C$_\alpha^3$C$_\alpha^8$, which were found to be most important in our model.
        For the TS, we report in transparency the superimposition of 20 configurations from the TSE.
        In the top row, we highlight the crucial hydrogen bonds between the Asp3, Thr6 and Thr8 residues, and for the TS, we report the structures related to the medoids of the two clusters.
        }
        \label{fig:chignolin}
    \end{figure*}
        
        But there is more, and in fact, the results of this analysis guided us to take a fresh look at this much-studied problem. 
        In fact, we noticed that a different way of expressing the alanine conformation can be obtained using as a descriptor the projection of the O coordinate on the direction perpendicular to the NCC$_\beta$ plane \enrico{(see SI, Sec.~\ref{sup_sec:o_projection})}.
        It turns out that this single descriptor approximates the reaction coordinate extremely well, reaching a very low variational minimum at $\mathcal{K}_m  = 1.2 $ a.u., which is comparable, within the statistical uncertainty, with the one obtained using the full interatomic distance set as descriptors.
        This exercise illustrates clearly one of the advantages of our approach, namely the ability of the method to focus the researcher's attention on where the real action is and to help design new efficient collective variables.
    
    \subsection{DASA reaction}\label{subsec:dasa}

        Our third test is the 4$\pi$-electrocyclization of the donor-acceptor Stenhouse adduct (DASA), which is depicted in Fig.~\ref{fig:dasa}, that is part of a complex photo-switching pathway~\cite{zulfikri2019dasa, stricker2022multi, raucci2022enhanced}. 
        This reaction involves a major conformational change, the formation of a cyclopentenone ring, and a proton transfer from $O_1$ to $O_2$~\cite{sanchez2021silico}. The energy barrier associated with such  transition is much larger than $k_BT$ ($\approx 1$ eV~\cite{raucci2022enhanced}), 
        thus, we expect the committor to have a very sharp step-like behavior.
        Although, in this case, it would probably have been more efficient to use standard methods based on the search for the stationary points of the potential energy surface, we found both instructive and challenging to solve the problem using our approach.
        Indeed, the iterative process needed to be adapted to the sharp features of the committor, as described in the method section.
    
        As we are dealing with a chemical reaction involving the formation and breaking of bonds, it is natural to take the distances between atoms involved in such bond modifications as descriptors.
        In particular, we select as input features the distances between atoms labeled in Fig.~\ref{fig:dasa}) for a total of 45 input descriptors. 
        
        After having obtained the committor, we find that four distances are ranked higher than the others (see SI, Fig.~\ref{sup_fig:dasa_rank}), which are, in the order, the C$_1$C$_5$, C$_1$O$_1$, C$_1$C$_6$, and C$_2$O$_2$ distances. 
        There are clear chemical reasons for this result.
        The formation of the C$_1$C$_5$ bond reflects the closure of the cyclopentenone ring, the three distances C$_1$O$_1$, C$_1$C$_6$, and C$_2$O$_2$ describe the conformational change that the molecule undergoes during the reaction and are operative in reducing the O$_1$O$_2$ distance, thus allowing the proton to be transferred from O$_1$ to O$_2$ (see SI, Fig.~\ref{sup_fig:dasa_distribution_dOO}). Finally, the role of the  C$_1$C$_6$ in the reaction reflects the change in bonding length due to the C$_5$ hybridization change from sp$^2$ to sp$^3$ that takes place when the pentagonal ring closes. 

        Contrary to the case of the Müller potential and of alanine dipeptide, already from a visual inspection, it is clear that the TSE exhibits some complexity (see SI, Fig.~\ref{sup_fig:dasa_TSall}).
        Thus, we apply the k-medoids analysis~\cite{Schubert2022kmedoids} and find that the TSE can be described as composed of two classes, which can be distinguished by the different puckering~\cite{cremer1975general} of the 1,3-dioxane hexagonal ring, as shown in Fig.~\ref{fig:dasa}.
        If optimized, the two structures differ in energy by only $\approx 1 k_BT$ thus, they both are likely to be part of the TSE. 
        This example, and more so the one that follows, demonstrates the need to introduce the TSE concept, thus extending the more traditional view of associating a single configuration to the transition state.
    
    \subsection{Chignolin}
        The final test of our method is the study of the chignolin protein in solution, which is able to fold in a stable hairpin structure. 
        Luckily, the unbiased 106-${\mu}s$-long trajectory performed by the D.E. Shaw group using a special purpose machine~\cite{lindorff2011fast} is available and provides a precious benchmark for our calculations.
        
        In this case, we use as descriptors all the 45 distances that connect the 10 $\alpha$-carbons as done in Ref.~\cite{ray2023deep}. 
        After reaching convergence \enrico{ of our procedure, by analyzing the 40-ns-long simulations of final iteration}, we collected a large number of TSE configurations that, at a visual inspection, appear to be very close to those obtained in the much longer unbiased dynamics of Ref.~\cite{lindorff2011fast}.  
        In fact, we find that in all TSE configurations, the hairpin bend ($4\,$-$\,7$) is formed, and the two prongs of the hairpin are roughly aligned.  
        However, the two protein segments ($1\,$-$\,4$) and ($7\,$-$\,10$) exhibit a variety of conformational arrangements, as depicted in Fig.~\ref{fig:chignolin}.
        Luckily, the large number of TSE configurations we were able to collect allowed a statistically meaningful analysis of the TSE to be carried out and the apparent disorder to be understood.     
      
        As in the previous cases, we perform a relevance analysis of the descriptors in the optimized  $q_\theta (\textbf{x})$ as in Ref.~\cite{bonati2020data}. 
        We find two surprising results: none of the descriptors that can be associated with the formation of the hairpin bend play a significant role. 
        Instead, the two apparently improbable distances C$_2$C$_6$ and C$_3$C$_8$  emerge as rather significant (see SI, Fig.~\ref{sup_fig:chignolin_rank}).  In a first instance, this result can be understood if we notice that the hairpin bend is easily formed, and many unfolded structures share this feature (see SI, Fig.~\ref{sup_fig:chignolin_distribution_d47}). 
        Furthermore, the role of the two distances can be attributed to the need to align the two prongs before folding, this being a significant and entropy-costly step on the way to folding. 
        But there is more to it, as we discovered by clustering the TSE data using again the k-medoids method~\cite{Schubert2022kmedoids}.  
        In fact, we find that the TSE configurations can be classified into two groups.  
        In one, a bidentate H-bond is established between  Asp3 and Thr6 (TSE$^{up}$). 
        In the other, a monodentate H-bond links Asp3 and Thr8 (TSE$^{down}$) (see also SI, Fig~\ref{sup_fig:chignolin_distribution_adjacentmatrix}).     
        We note that in the folded state, both H-bonds are simultaneously formed, while in the unfolded state, the probability of finding either of these two bonds formed is very low (see SI,  Fig.~\ref{sup_fig:chignolin_distribution_adjacentmatrix}). 
        
        Since the formation of these H-bonds was not explicitly included in the descriptors, the two distances 
        C$_2$C$_6$ and C$_3$C$_8$ also act as proxies for the formation on the way to the folded state of either one of these two crucial H-bonds~\cite{bonati2021deep}.        
        
\section{Discussion}
    In this paper, we have developed a new strategy for tackling the rare event problem. 
    This strategy is guided by Kolmogorov's variational principle for the determination of the committor function and has led us to define what we call the Kolmogorov ensemble, in which a committor-dependent bias is added to the interatomic potential.  
    This natural extension of the notion of transition state is indispensable to describe complex systems, like chignolin, DASA, and many other~\cite{bonati2023catalysis}, where it is not possible to identify a single state as the one through which the reaction has to pass.  
    We have also shown that the variational principle provides a powerful tool for analyzing the TSE, identifying the most relevant degrees of freedom involved in the reaction, and ranking them in a quantitative way. 
   
    This and the availability of a large number of TSE configurations can help construct efficient collective variables~\cite{ray2023deep} both for enhancing sampling and summarizing the physics of the process under study. 
    Knowledge of the TSE will also be extremely useful when building reactive machine learning potentials, where collecting data on the transition state has proven to be essential for obtaining reliable results.~\cite{yang2022using} 

    However, the most exciting perspective is that from the sampling of the TSE, and with the help of the analysis tools that we have developed, we can gain new and deep insight into reactive processes so as to be able to unveil enzymatic reaction mechanisms, steer a chemical reaction towards a desired product, design new drugs and even guide crystallization processes.

 

\section{Methods}

    \subsection{Self-consistent iterative procedure}
        As anticipated in the introduction, we represent the committor function $q(\textbf{x})$ as the output of a neural network (NN) $q_\theta(\textbf{d}(\textbf{x}))$, with $\theta$ trainable parameters, which takes as input a set of (physical) descriptors $\textbf{d}(\textbf{x})$ \enrico{that are functions of the atomic coordinates $\textbf{x}$}.
        We optimize our model using the variational principle of Eq.~\ref{eq:variational_functional}. To do so, we use statistical sampling to evaluate the integral in $\mathcal{K}[q(\textbf{x})]$ and rely on a self-consistent iterative procedure in which we alternate cycles of training to cycles of sampling.
        This section briefly presents only a schematic overview of the steps involved in such a procedure, discussing the details of the different components in the following sections.
    
        \begin{itemize}
            \item \textbf{Step 1:}  The committor   $q_\theta^n(\textbf{x})$ at iteration n is constructed  using the dataset of configurations $\mathbf{x}^{n}$ and weights $w_i^{n}$ updated from the previous iterations (see Sec.~\ref{subsec:optimization}). 
            For the first iteration $n=0$, we shall use a dataset consisting of configurations collected with unbiased simulations in the two metastable basins and labeled accordingly.
            
            \item \textbf{Step 2:} We perform biased simulations to sample the \emph{Kolmogorov ensemble} defined in Eq.~\ref{eq:kolmogorov_distribution}. These simulations can be started from the two basins or even from the TSE.
            We do this  by applying the bias  $V_\mathcal{K}^n(\textbf{x}) = -\frac{1}{\beta} \log(|\nabla q^{n}_\theta(\textbf{x})|^2)$ (see Sec.~\ref{subsec:sampling}). 
            We check whether convergence has been reached. If not, we proceed to step 3.
            \item \textbf{Step 3:} We update our training set with the new sampled configurations, reweighing them by the applied bias $w^n_i = \frac{e^{\beta V_\mathcal{K}^n(\textbf{x}_i)}}{\langle e^{\beta V_\mathcal{K}^n(\textbf{x})} \rangle_{U^n_\mathcal{K}} }$ (see Sec.~\ref{subsec:sampling}) and repeat from step 1.
        \end{itemize}

    \subsection{ Optimization strategy }\label{subsec:optimization}
        To optimize the committor model, we translate the Kolmogorov variational functional and the related boundary conditions into a loss function composed of two terms.
    
        A key \emph{variational loss} term $L_v$, that is used to evaluate the functional in Eq.~\ref{eq:variational_functional}, as
            \begin{equation}
                L_v = \frac{1}{N^n} \sum_i^{N^n}  w_i |\nabla_\textbf{u} q(\textbf{x}_i)|^2
                \label{eq:loss_variational}
            \end{equation}
        where we use as training set  all the $N^n$ configurations  $\textbf{x}_i$  collected \enrico{until} iteration $n$ together with their associated statistical weights  $w_i$  and $\nabla_\textbf{u}$ denotes the gradient with respect to the mass-weighted coordinates $\textbf{u}_i^j = \sqrt{m^j} \textbf{x}_i^j$, in which $m^j$ is the mass of atoms of type $j$.
        
        This term is complemented by the \emph{boundary loss} term $L_b$, which imposes the correct boundary conditions, i.e., $q(\textbf{x}_A)=0$ and $q(\textbf{x}_B)=1$, and is expressed as
            \begin{equation}
                L_b = \frac{1}{N_A}\sum_{i \in A}^{N_A} (q(\textbf{x}_i))^2 + \frac{1}{N_B}\sum_{i \in B}^{N_B} (q(\textbf{x}_i) - 1)^2
                \label{eq:loss_boundary}
            \end{equation}
        This term is computed only on the labeled dataset introduced in the first iteration $n=0$ that consists of $N_A$ unbiased configurations from state A and $N_B$ configurations from state B.

        The total loss function is thus obtained as a linear combination of these components
            \begin{equation}
                L = L_v + \alpha L_b
                \label{eq:total_loss}
            \end{equation}
        in which we introduce the $\alpha$ hyperparameter to scale the relative contributions of the two terms during the optimization procedure.
        It is worth noting that, in the first iteration, the $L_v$ contribution to the total loss will be minimal, as the dataset is limited to close-to-equilibrium configurations from the bottom of metastable states. 
        Nonetheless, the $L_b$ term still allows obtaining a reasonable first guess $q^0_\theta(\textbf{x})$, which can be seen as a classifier trained to distinguish between states A and B. 
        This is not surprising considering that a very similar approach has often been used to design machine learning collective variables for enhanced sampling.~\cite{bonati2020data, trizio2021enhanced}

    \subsection{Sampling the Kolmogorov ensemble}\label{subsec:sampling}
        As discussed in the introduction, the variational approach of Eq.~\ref{eq:variational_functional} and the corresponding loss term of Eq.~\ref{eq:loss_variational} are of little use if the TSE, where $|\nabla q(\textbf{x})|^2$ is significantly different from zero, is poorly represented in the training dataset.
        
        In previous applications of the Kolmogorov variational principle, enhanced sampling methods were used to collect data in the TS region.~\cite{li2019computing, rotskoff2020learning} 
        However, they relied on the use of collective variables (CVs), and even assuming that the CV is able to capture the TSE main features, they had to spend time sampling over and over uninteresting regions of the configuration space, such as those belonging to the metastable states. 
                
        In contrast, we apply to the system a bias that is attractive in the TS region and repulsive in the basin regions. 
        Even using a simplified model for $q_\theta(\textbf{x})$ the addition of the potential        
            \begin{equation}
                V_\mathcal{K}(\textbf{x}) = -\frac{1}{\beta} \log(|\nabla q_\theta(\textbf{x})|^2 + \epsilon)
                \label{eq:bias}
            \end{equation}
        where  $\epsilon$ is a positive regularization term, biases the sampling towards the TS and away from basin $A$ and $B$.
        Thus, already after the first iterations, TSE configurations are being harvested and attention is taken away from the metastable basins. \enrico{It should be noted that in practice, the bias in Eq.~\ref{eq:bias} can be computed by the gradient with respect to the input features of $q_\theta$ for a simpler and faster interface with PLUMED~\cite{plumed2019promoting} (see Sec.~\ref{sec:codes_software}).  At convergence, the results will not depend on this choice since, when computing $\mathcal {K}[q(\textbf x)]$, the configurations thus generated are reweighed to give them the correct Boltzmann weight.}
        This is done by associating each configuration $i$ that was added to the training set at iteration $n$ with a weight 
            \begin{equation}
                w^n_i = \frac{e^{\beta V_\mathcal{K}^n(\textbf{x}_i)}}{\langle e^{\beta V_\mathcal{K}^n(\textbf{x})} \rangle_{U_\mathcal{K}^n} }
                \label{eq:weights}
            \end{equation}
      which does not explode exponentially given the logarithmic nature of the bias Eq.~\ref{eq:bias}.

    \subsection{Tips and tricks for optimization}
        In our experience, the straightforward version of the iterative method presented above leads to convergence after an affordable number of iterations.
        However, to accelerate convergence, it is expedient to introduce some modification to the self-consistent cycle guided, if possible, by previous qualitative knowledge of the system.

        For example, as the configurations from the first iteration labeled dataset are unbiased, they are assigned unitary weights $w^0_{i \in A} = w^0_{i \in B} = \exp(\beta V_\mathcal{K}^0(\textbf{x}_i)) = 1$. This implicitly implies that in the Boltzmann ensemble, the two metastable states have the same energy and can be sampled with the same probability.
        In spite of this unphysical assumption, after a few iterations, the correct relative statistical weight between the points in the two basins is re-established.  
        However, the number of iterations needed to reach convergence can be reduced if we have an even approximate estimate of the free energy difference $\Delta F_{AB}$ between the initial and final basin. 
        In such a case, we can use a less approximate dataset on which we modify the weights of the initial points in $B$ as
            \begin{equation}
                w^0_{i\in B} = 1 \quad \rightarrow \quad \tilde{w}_{i\in B} = \exp(-\beta \Delta F_{AB})
            \end{equation}
        to make the underlying distribution resemble more closely the Boltzmann one (see SI, Fig.~\ref{sup_fig:muller_deltaf_comparison}).

        Of course, if we have other information on the TSE coming, for instance, from enhanced sampling simulations or successful molecular dynamics runs in which reactive trajectories have been obtained, we can use them from the initial iteration to obtain a better starting guess and a speedier convergence.
        
        Since the bias is used here to speed up the calculation, we are at liberty to change its magnitude, provided that the data collected are properly reweighed.  
        In this respect, the simplest and most controllable device is to multiply the bias in Eq.~\ref{eq:bias} by a positive multiplicative factor $\lambda$
            \begin{equation}
                V_K(\textbf{x}) \quad \rightarrow \quad 
                \tilde{V}_K(\textbf{x}) = \lambda V_K(\textbf{x})
            \end{equation}
        In the first iteration, we run several parallel simulations with different values  $\lambda \sim 1$, choose among the $\lambda$ values tested the smallest that is capable of attracting the system to the TSE, and in the following iterations, we keep $\lambda$ fixed to this value. 
        In the case of DASA, that is representative of systems in which the committor changes very quickly in a small region of configuration space such that $|\nabla q(x)|^2$ can assume large values, the bias can become too large and trap the system in the TSE.  
        To remedy this problem, from the second iteration, we explore the effect of a range of $\lambda \sim 1$ values and this time, we choose the largest value of $\lambda$ that allows escaping the TSE and we maintain this value in the following iteration. In hard cases, one can further optimize the value of $\lambda$ at each iteration, varying $\lambda$ in a small range of values and picking again the one that is most effective. 
        It must be added that the computer time invested in the simulations needed to improve the choice of $\lambda$ is not wasted, as the configurations and statistical weights thus collected can be added to the training set and thus used to improve our estimation of the integral in $\mathcal{K}[q(\textbf{x})]$.
       
         To avoid an artificial bias, the data needed to pass at successive iterations are collected by combining data from simulations that start from both $A$ and $B$.  
         However, if, as in the case of the DASA, $|\nabla q(\textbf{x})|^2$ is strongly peaked, once the TSE has been visited, it is helpful to start the successive iterations also from  TSE configurations. 
         This is because when the committor has a sharp step-like behavior, the action of the bias will be confined to a very narrow region, eventually making it difficult for simulations that start from either $A$ or $B$ to reach the TSE region. 

         \enrico{Based on the variational nature of the optimization criterion, the $\mathcal{K}_m$ value, as estimated in practice in Eq.~\ref{eq:loss_variational}, typically suffices as a figure of merit for monitoring the convergence of the procedure and the accuracy of the obtained model. 
         However, if desired, further evaluation could be performed by training a committee of models to obtain a statistical measure of the model uncertainty, similarly to what is commonly done when dealing with machine learning potentials.~\cite{yang2022using} }

    \subsection{Codes and software}
        \label{sec:codes_software}
        The reported NN-based committor models are based on the Python machine learning library PyTorch~\cite{paszke2019pytorch}. 
        The specific code for the definition and the training of the model is developed in the framework of the open-source \texttt{mlcolvar}~\cite{bonati2023mlcolvar} library. 
        The committor-based enhanced sampling simulations have been performed using the open-source plugin PLUMED~\cite{tribello2014plumed} 2.9, modifying the \texttt{PYTORCH\_MODEL} interface available in the optional \texttt{pytorch}~\cite{bonati2023mlcolvar} module of the code.
        This has been patched with different MD engines to simulate the reported systems. 
        The M\"uller-Brown potential's simulations have been performed using the MD engine in the \texttt{ves\_md\_linearexpansion}~\cite{valsson2014variational} module of PLUMED.
        The vacuum alanine dipeptide simulations have been carried out using the GROMACS v2021.5~\cite{abraham2015gromacs} MD engine and the Amber99-SB~\cite{amber2013} force field.
        The DASA reaction simulations have been carried out using the CP2K-8.1~\cite{cp2k2020} software package at PM6 semi-empirical level\cite{stewart2007optimization}.
        For the study of folding and unfolding of chignolin in explicit solvent, we performed our simulations using GROMACS v2021.5~\cite{abraham2015gromacs} the CHARMM22$^*$~\cite{piana2011robust} force field and TIP3P~\cite{mackerell1998tip} water force field.
        All the reported molecular snapshots have been produced using the open-source PyMOL~\cite{PyMOL} code, whereas the clustering analyses have been performed using the k-medoids method as implemented in the \texttt{kmedoids}~\cite{Schubert2022kmedoids} Python library.
        


%
%

%

\begin{acknowledgments}
The authors want to acknowledge Umberto Raucci for the suggestions about the DASA, Dhiman Ray for the hints on clustering methods, Andrea Rizzi and Luigi Bonati for code design contributions, and Francesco Mambretti for the many helpful conversations and for carefully reading the manuscript.

\end{acknowledgments}

\section*{Code and Data Availability} \label{sec:code_avail}
    The code for the definition and training of NN-based committor model will be released through the \texttt{mlcolvar} library~\cite{bonati2023mlcolvar} upon publication.
    Similarly, the PLUMED inputs and bias interface will be made available on the PLUMED-NEST~\cite{plumed2019promoting} repository. 

\section*{Bibliography}
\bibliography{references}

\setcounter{section}{0}
\renewcommand{\thesection}{S\arabic{section}}
\setcounter{equation}{0}
\renewcommand{\theequation}{S\arabic{equation}}
\setcounter{figure}{0}
\renewcommand{\thefigure}{S\arabic{figure}}
\setcounter{table}{0}
\renewcommand{\thetable}{S\arabic{table}}
    
\clearpage
\onecolumngrid

{\Large\normalfont\sffamily\bfseries{{Supporting Information}}}

\setlength{\tabcolsep}{18pt}
\renewcommand{\arraystretch}{1.2}

\begin{figure*}[h!]
        \centering
        \begin{minipage}{0.3\linewidth}
           \includegraphics[width=\linewidth]{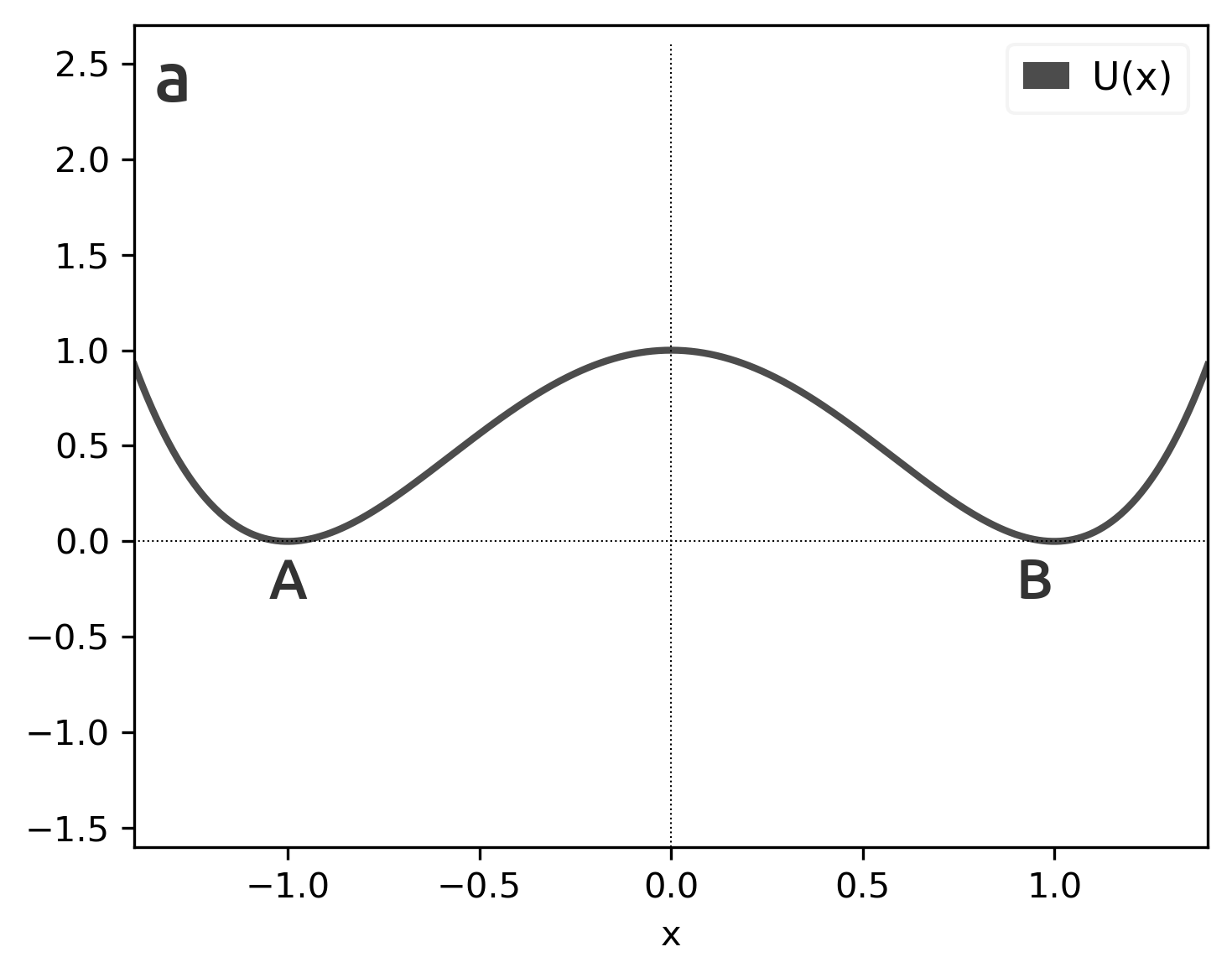}
        \end{minipage}
        \begin{minipage}{0.3\linewidth}
           \includegraphics[width=\linewidth]{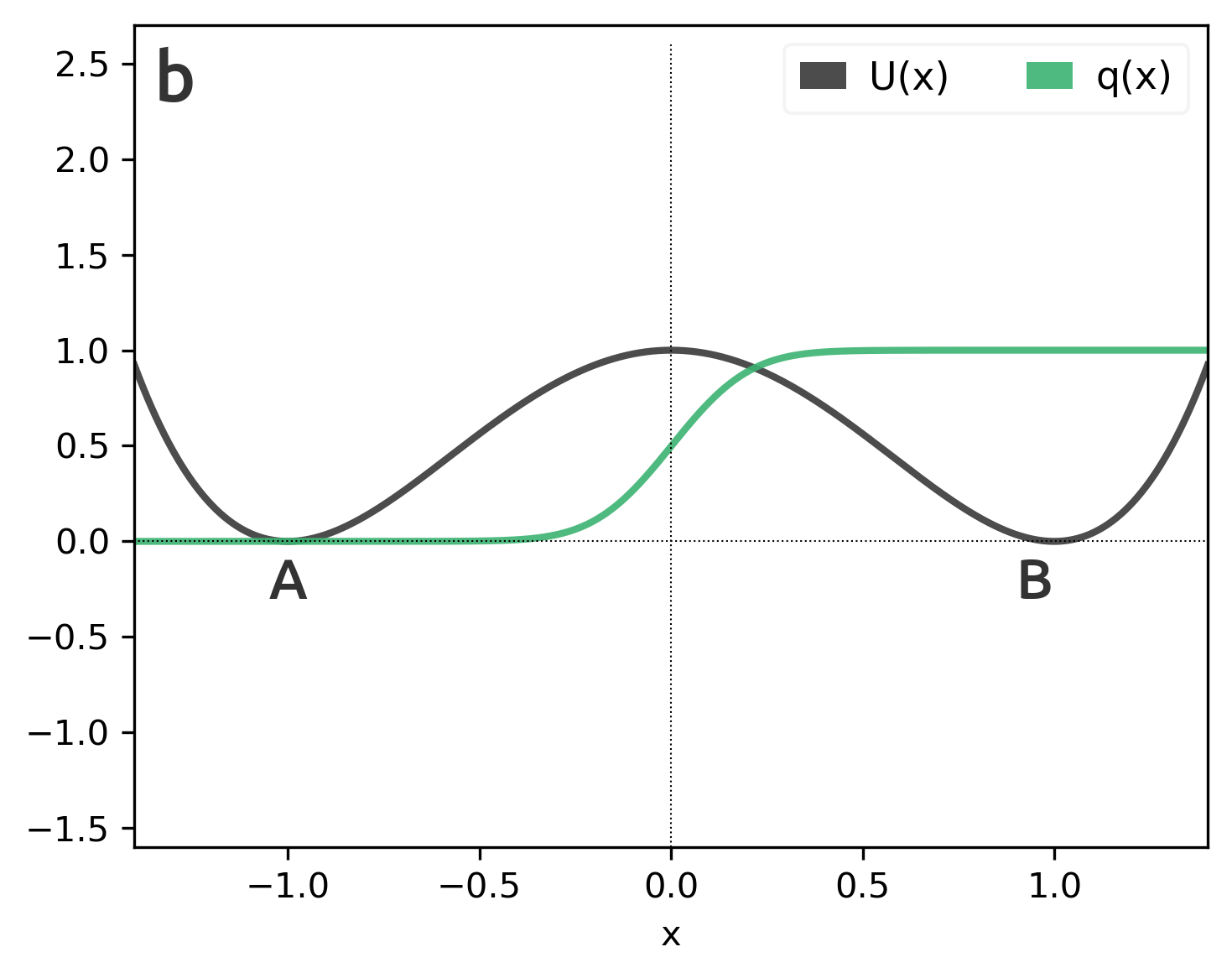} 
        \end{minipage}
        \begin{minipage}{0.3\linewidth}
           \includegraphics[width=\linewidth]{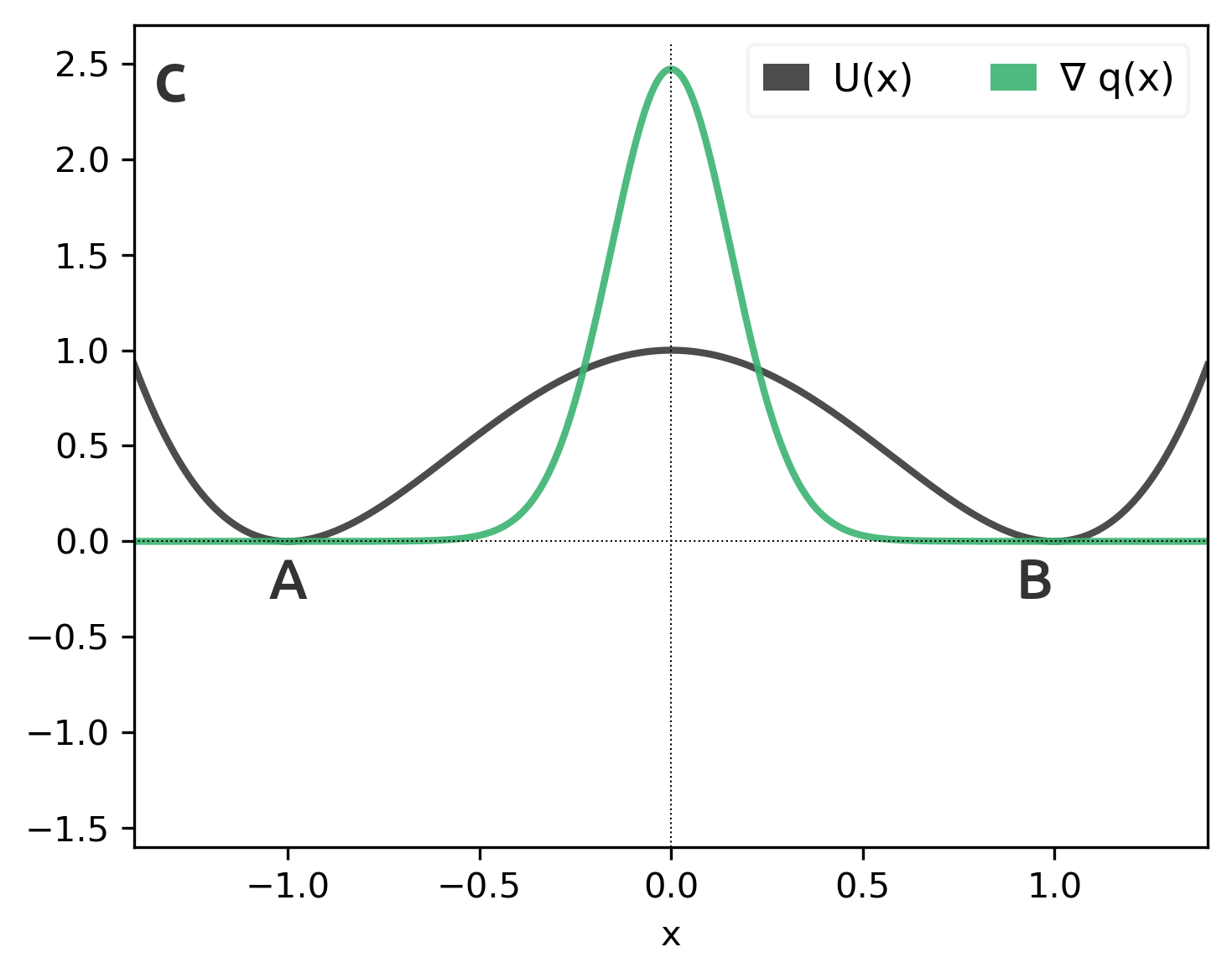}
        \end{minipage}
        
        \begin{minipage}{0.3\linewidth}
           \includegraphics[width=\linewidth]{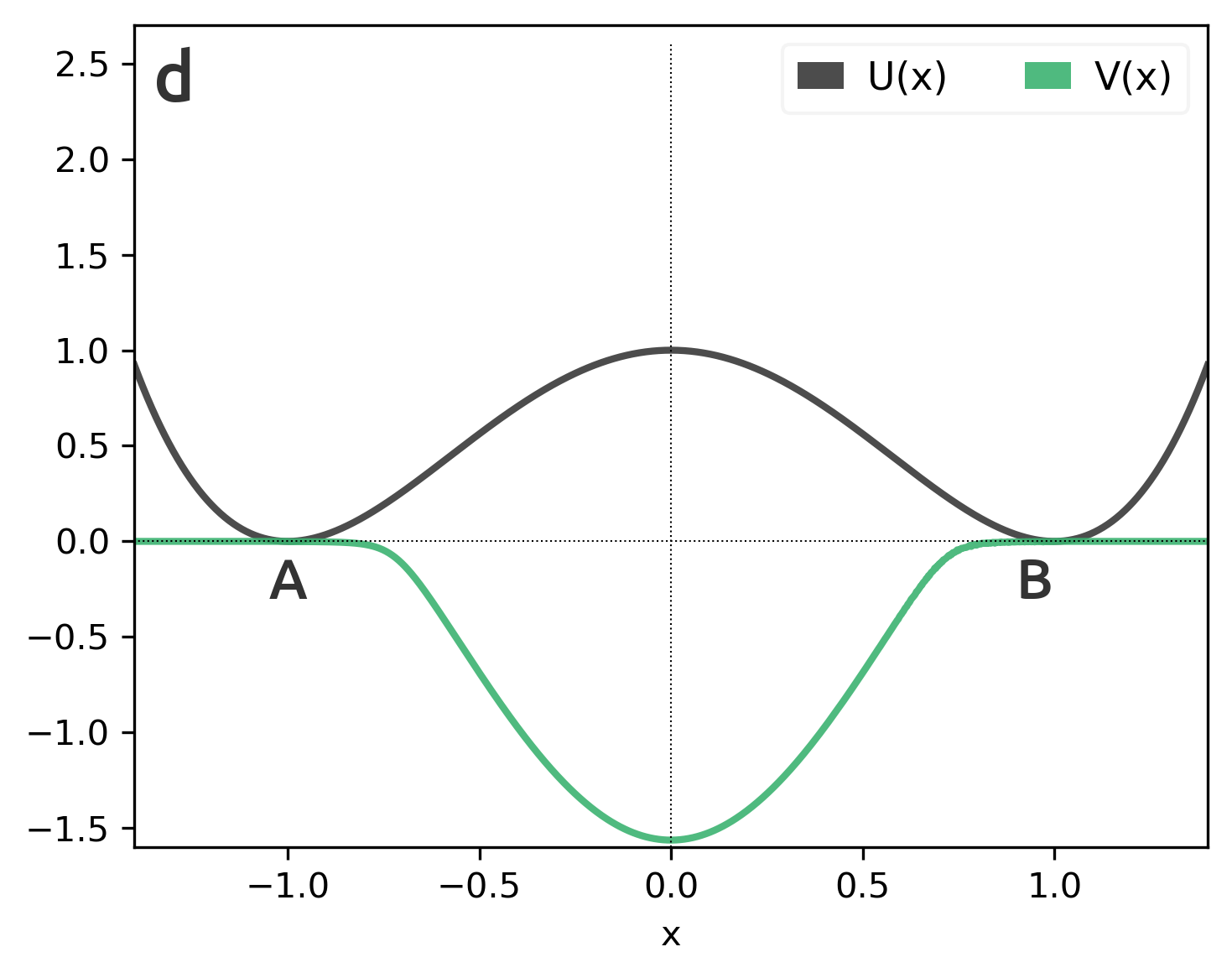} 
        \end{minipage}
        \begin{minipage}{0.3\linewidth}
           \includegraphics[width=\linewidth]{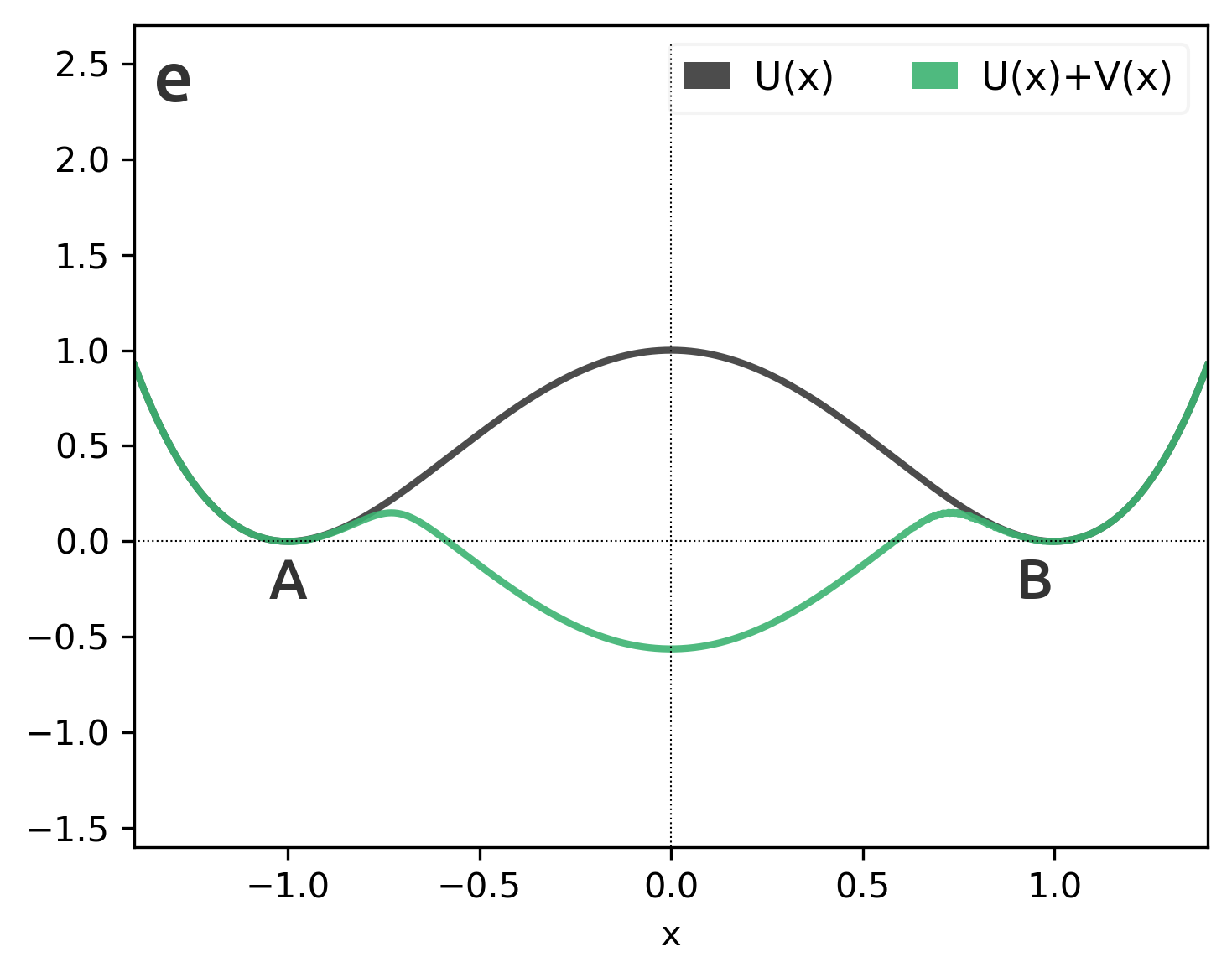}
        \end{minipage}
        \caption{\enrico{Schematic visualization of committor-based bias potential for TSE sampling on a toy double-well potential. \textbf{a)} Potential energy $U(x)$ (repeated in all other panels in black) \textbf{b)} Committor function $q(x)$ \textbf{c)} Gradient of committor function $\nabla q(x)$ \textbf{d)} Committor-based bias potential $V(x)$ \textbf{e)} Biased energy landscape $U_\mathcal{K} = U(x) + V(x)$ for extensive TSE sampling.}}
        \label{sup_fig:visual_bias}
    \end{figure*}

\enrico{\section{Short note on Eq.~\ref{eq:variational_functional}}
    \label{sup_sec:variational_principle}
    The committor function $q(\textbf{x})$ from $A$ to $B$ over a domain $\Omega$, under the hypothesis of overdamped dynamics, can be obtained as the solution of a set of partial differential equation~~\cite{weinan2010transition}
        \begin{equation}
            \begin{cases}
                \nabla U \cdot \nabla q - \beta^{-1} \Delta q = 0 \qquad \textbf{x} \in \Omega\setminus(A\cup B) \\
                q(\textbf{x}) = 0 \qquad \textbf{x} \in A \\
                q(\textbf{x}) = 1 \qquad \textbf{x} \in B \\
            \end{cases}
            \label{seq:kolmogorov_PDE}
        \end{equation}
    which, unfortunately, can only be solved for extremely simple toy models, such as the toy double well potential reported in Fig.~\ref{sup_fig:visual_bias}. 
    However, an equivalent solution can also be obtained by minimization of the variational functional $\mathcal{K}$~\cite{weinan2010transition} 
        \begin{equation}
            \begin{cases}
                \displaystyle \min_q \mathcal{K}[q(\textbf{x})] : \mathcal{K}[q(\textbf{x})] = \frac{1}{\mathcal{Z}} \int |\nabla q(\textbf{x})|^2 e^{-\beta U(\textbf{x})}d\textbf{x} \qquad
                \textbf{x} \in \Omega\setminus(A\cup B) \\
                q(\textbf{x}) = 0 \qquad \textbf{x} \in A \\
                q(\textbf{x}) = 1 \qquad \textbf{x} \in B \\
            \end{cases}
            \label{seq:kolmogorov_VAR}
        \end{equation}
    where $U(\textbf{x})$ is the interatomic potential and $\mathcal{Z}=\int e^{-\beta U (\textbf{x})} d\textbf{x}$ is the corresponding partition function. \\
    We then observe that the variational functional of Eq.~\ref{seq:kolmogorov_VAR} can also be written as the ensemble average of the $|\nabla q(\textbf{x})|^2$ quantity over the Boltzmann ensemble driven by the potential $U(\textbf{x})$, as reported in Eq.~\ref{eq:variational_functional} in the main text.
        \begin{equation}
            \mathcal{K}[q(\textbf{x})] \quad=\quad 
            \frac{1}{\mathcal{Z}} \int |\nabla q(\textbf{x})|^2 e^{-\beta U(\textbf{x})}d\textbf{x} \quad=\quad
            \Big \langle \big|\nabla q(\textbf{x})\big|^2 \Big\rangle_{U(\textbf{x})}
            \label{sup_eq:variational_full}
        \end{equation}
    Despite being derived under the overdamped dynamics hypothesis, this whole formalism can also be extended to the general case of Langevin equation by introducing a few reasonable approximations, as discussed in detail in Sec.~3.5 of Ref.~\cite{weinan2010transition}
}

        \begin{figure}[h!]
            \centering
            \includegraphics[width=0.5
            \linewidth]{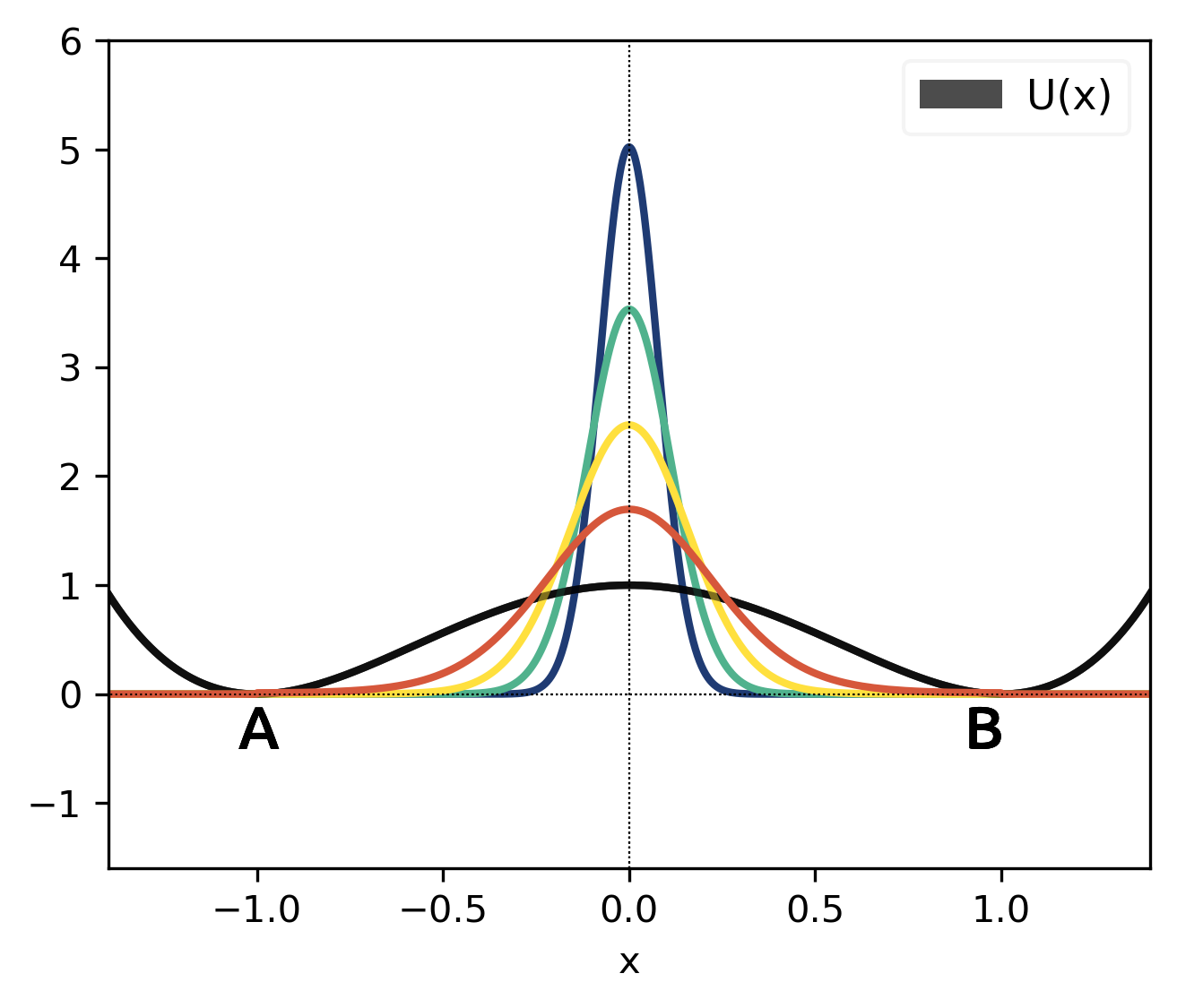}
            \caption{\enrico{Normalized contribution to the integral of the variational function reported in Eq.~\ref{sup_eq:variational_full} in the case of a toy double-well potential (reported in black) for different $\beta$ values [5, 10, 20, 40]. Colder colors correspond to lower temperatures, and warmer colors to higher temperatures.}}
            \label{sup_fig:var_integral_contribution}
        \end{figure}

\enrico{\section{Visualization of committor-based bias potential}
    \label{sup_sec:bias_motivation}
    In order to apply the variational principle of Eq.~\ref{eq:variational_functional} (see also Sec.~\ref{sup_sec:variational_principle}), extensive sampling of the TS region is needed, as the relevant contributions to the $\mathcal{K}[q(\textbf{x})]$ functional come from that region.
    To collect such data, it would be ideal to have a bias potential for an enhanced sampling simulation that can focus sampling toward the TS region.
    However, the determination of such a bias appears to be a chicken-and-egg problem. Indeed, in order to build such bias, one would need to be able to know the TS in advance, and in order to know the TS, one would need to have such a bias. \\
    It is known~\cite{weinan2010transition} that the committor function $q(\textbf{x})$ provides a way to mathematically formalize the concept of TS. 
    Conventionally, this is localized to the region where $q(\textbf{x})\simeq\frac{1}{2}$, as schematically depicted in the case of a toy double-well potential in panel b of Fig.~\ref{sup_fig:visual_bias}.
    It is interesting to note that, as a consequence, the gradients $\nabla q(\textbf{x})$ of the committor are localized on the TSE region (see panel c and also Fig.~\ref{sup_fig:var_integral_contribution}). \\
    This peculiar property of $q(\textbf{x})$ motivated us to formulate our TS-oriented bias potential (see Eq.~\ref{eq:bias} and panel d) as a function of $|\nabla q(\textbf{x})|^2$, guaranteeing its focus on the TS region by design and thus providing a rather simple solution out of the aforementioned chicken-and-egg problem. We note that, in Eq.~\ref{eq:bias}, we introduce the logarithm to have a smoother behavior of the bias and easier reweighing.
    Indeed, when applied to the system, such bias can transform the TS into a minimum that can be effortlessly and extensively sampled (see panel e).
}

\begin{figure}[h!]
            \centering
            \includegraphics[width=0.55
            \linewidth]{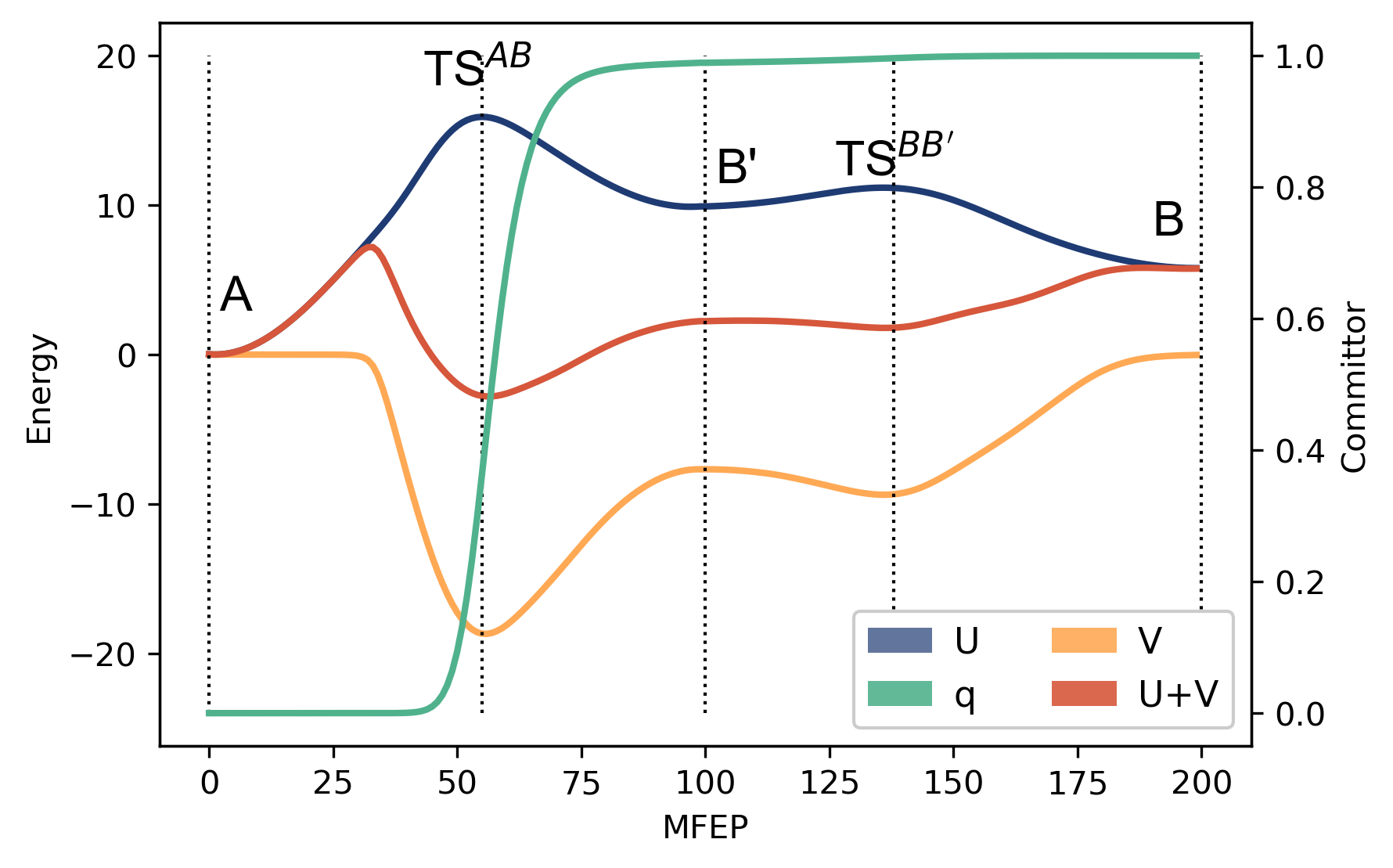}
            \caption{\enrico{Projection along the minimum free energy path (MFEP) of the Muller potential energy (U), committor function (q), committor-based bias (V), and biased energy landscape (U+V). }}
            \label{sup_fig:muller_mfep}
        \end{figure}

    \enrico{It is also instructive to visualize the effect of the bias potential on the M\"uller potential along the minimum free energy path (MFEP) connecting basins A and B passing through the TS, which we report in Fig.~\ref{sup_fig:muller_mfep}. Along the MFEP, the A and B basins can be clearly distinguished, as well as the state we label as B', which can be seen as a model example of a shallow intermediate state. It can be seen how the committor-based bias potential reflects the features of the underlying potential energy landscape, focusing its action on the region associated with the barriers.}

\begin{figure}[h!]
            \centering
            \includegraphics[width=0.45
            \linewidth]{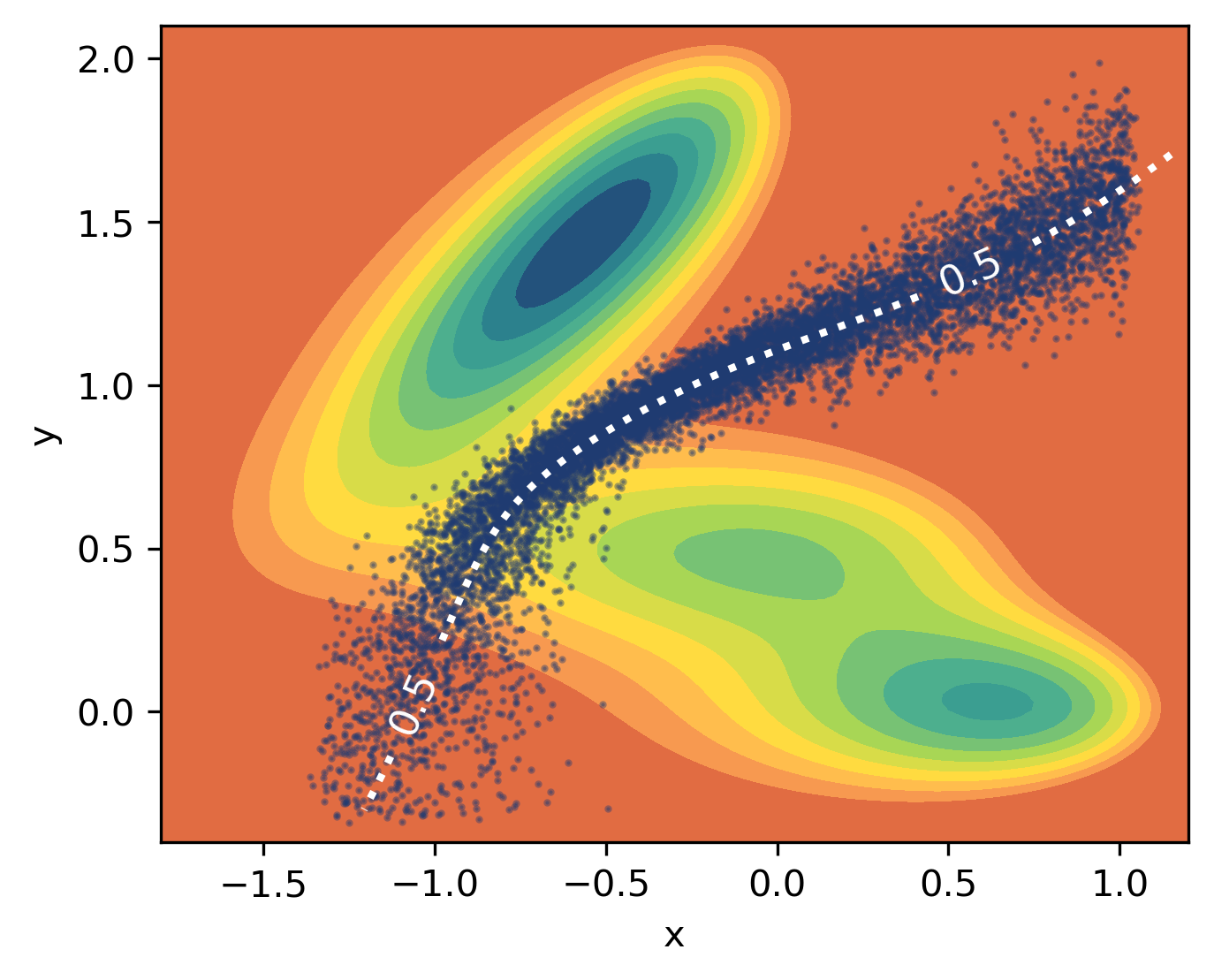}
            \caption{\enrico{Scatter plot of the sampling using only $V_\mathcal{K}$ as an effective potential on M\"uller potential energy surface compared with the isocommittor line $q = 0.5$, reported as a white dotted line.}}
            \label{sup_fig:muller_pv_sampling}
        \end{figure}

    \enrico{In the main text, we also link the sampling under the action of our bias potential to the conventional $q\simeq\frac{1}{2}$ criterion based. 
    In Fig.~\ref{sup_fig:muller_pv_sampling}, we report the sampling under the action of $V_\mathcal{K}$ only in the case of the M\"uller system. Such a result is compared with the 0.5 isoline of the numerical reference of the committor function. 
    The correspondence of the two criteria can also be appreciated in Fig.~\ref{sup_fig:muller_mfep}}.

\newpage
\section{Ranking of descriptors with feature relevance analysis}\label{sup_sec:feature_ranking}
    To identify the most relevant inputs in our learned committor models, we rank the input descriptors by performing a feature relevance analysis.
    This is based on the derivatives of the committor model $q_\theta(\textbf{d}(\textbf{x}))$ with respect to the descriptors $\textbf{d}(\textbf{x})$ and the rank $r^k$ of descriptor $k$ is defined as
        \begin{equation}
            r^{k} =\sum_{x_{i} \in TSE} { \left| \frac{\partial q}{\partial d_{j}^{k}}\right|} \sigma(d^k)
        \end{equation}
     where the sum is performed over a set of TSE configurations defined by $q$ value (0.4<$q$<0.6), and $\sigma(d^k)$ is the standard deviation of descriptor $d^k$ over this set.
     We must note, however, that different approaches to feature relevance analysis in neural networks are available in specific machine learning literature~\cite{pizarroso2022sensitivity, novelli2022lasso}. 

\section{Choice of neural network activation functions}
    Considering that in our approach, the derivatives of the output of the neural network model are as important as the output itself, we used as an activation function for the hidden layers the hyperbolic tangent function (\texttt{tanh}). 
    This indeed provides a good trade-off between non-linear contribution to the model and guaranteeing stable and smooth derivatives~\cite{khoo2019solving}.
    
    Moreover, to facilitate the learning of a $q_\theta(\textbf{x})$ with the correct shape, we used a sharp sigmoid-like activation $s(y)$ for the last layer
    \begin{equation}
        s(y) = \frac{1}{1 + e^{-3y}}
    \end{equation}  

\section{Müller-Brown Potential - Additional information}
    \subsection{Computational details}
        \subsubsection*{Simulations details}
            The M\"uller-Brown potential energy surface, U(x, y), is defined as a function of the Cartesian coordinates x and y 
                \begin{equation}
                    U(x,y) = -k\sum_{i=1}^4 d_i e^{a_i(x-x_i) + b_i (x-x_i)(y-y_i) + c_i(y-y_i)}  
                    \label{seq:muller_formula}
                \end{equation}
            where the constants take the following values, k = 0.15, [$d_1$, $d_2$,  $d_3$,  $d_4$ ]= [-200, -100, -170, 15], [$a_1$, $a_2$,  $a_3$,  $a_4$] = [-1, -1, -6.5, 0.7], [$b_1$, $b_2$,  $b_3$,  $b_4$] = [0, 0, 11, 0.6], [$c_1$, $c_2$,  $c_3$,  $c_4$] = [-10, -10, -6.5, 0.7], [$x_1$, $x_2$,  $x_3$,  $x_4$] = [1, 0, -0.5, -1] and  [$y_1$, $y_2$,  $y_3$,  $y_4$]  = [0, 0.5, 1.5, 1].
            
            The simulations of the diffusion of an ideal particle of mass 1 have been performed using Langevin dynamics based on the Bussi-Parrinello algorithm~\cite{bussi2007accurate} as implemented in the \texttt{ves\_md\_linearexpansion}~\cite{valsson2014variational} module of PLUMED.
            The damping constant in the Langevin equation was set to  10/time-unit. 
            The time unit was defined arbitrarily and corresponds to 200 timesteps and natural units ($k_BT = 1$) were used in all the calculations.

        \subsubsection*{Committor model training details}
            To model the committor function $q_\theta(\textbf{x})$ at each iteration, we used the x and y Cartesian coordinates of the diffusing particle as inputs of a neural network (NN) with architecture [2, 20, 20, 1] nodes/layer.
            For the optimization, we used the ADAM optimizer with an initial learning rate of $10^{-3}$ modulated by an exponential decay with multiplicative factor $\gamma=0.99999$. The training was performed for $\sim$20000 epochs. The $\alpha$ hyperparameter in Eq.~\ref{eq:total_loss} was set to 10.
            The number of iterations, the corresponding dataset size, and the $\lambda$ used in the biased simulations are summarized in Table~\ref{sup_tab:muller_iterations} alongside the lowest value obtained for functional $K_m$ (e.q. the variational loss term $L_v$ ), which provides a quality and convergence measure\enrico{, the simulation time $t_s$ and the output sampling time $t_o$. To have a direct comparison with the reference numerical result $\mathcal{K}_m=4.18$, the reported $\mathcal{K}_m$ values are computed on the ideal dataset described in Sec.~\ref{sup_sec:numerical_muller}}.
        \begin{table}[h!]
            \caption {Summary of iterative procedure for M\"uller-Brown potential.} \label{sup_tab:muller_iterations}
            \begin{center}
            
            \begin{tabular}{ |c|c|c|c|c|c| } 
             \hline
             Iteration & Dataset size & $K_m$ & $\lambda$ & \enrico{$t_s$ [a.u.]} & \enrico{$t_o$ [a.u.]} \\ 
             \hline
                0   & 4000 &  77.8 & -   & \enrico{2*400000} & \enrico{200} \\
                1   & 24000 &  4.71 & 1  & \enrico{2*500000} & \enrico{50} \\
                2   & 44000 &  4.52 & 1  & \enrico{2*500000} & \enrico{50} \\
                3   & 64000 &  4.48 & 1  & \enrico{2*500000} & \enrico{50} \\[1ex] 
             \hline
            \end{tabular}
            \end{center}
        \end{table}

\begin{figure}[h!]
            \centering
            \includegraphics[width=0.4\linewidth]{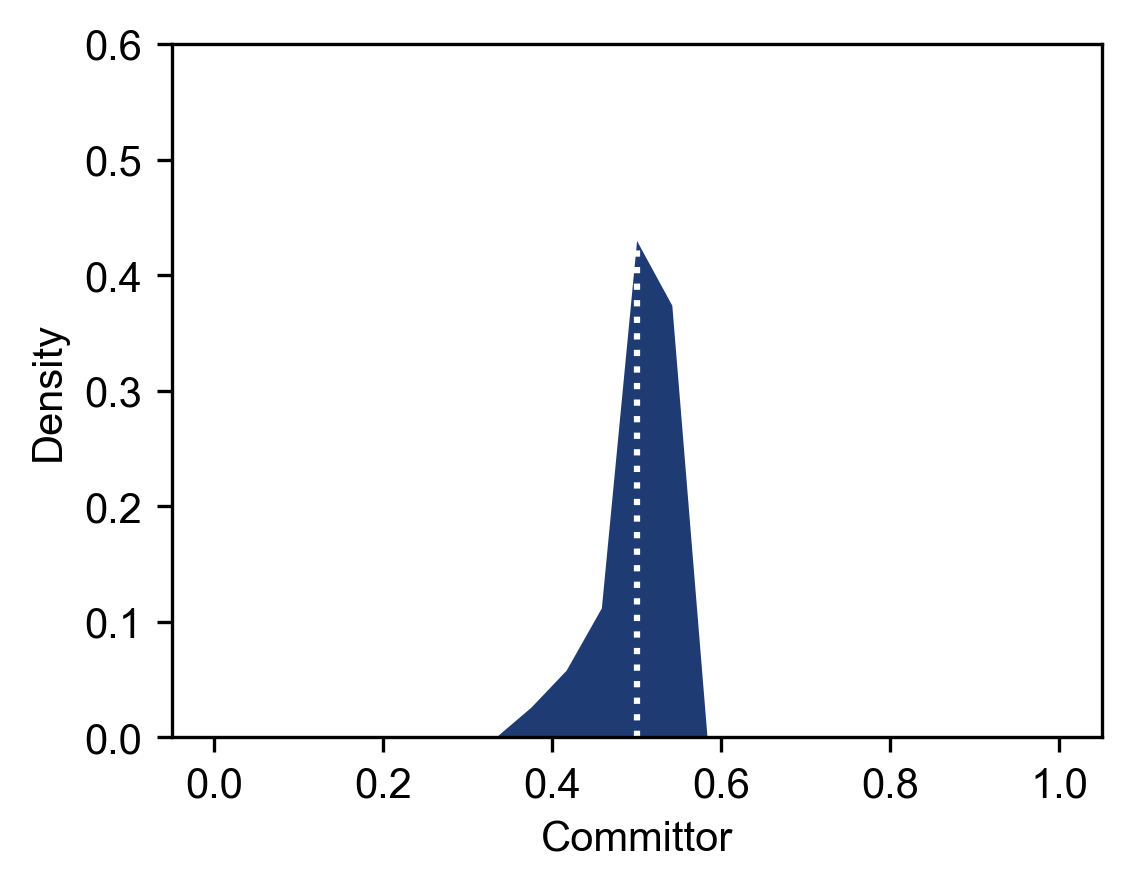}    
            \caption{\enrico{Normalized distribution of the results of the committor analysis for a set of 500 M\"uller-Brown configurations with $0.45<q<0.55$ sampled with our approach.}}
            \label{sup_fig:muller_committor_analysis}
        \end{figure}

    \enrico{In Fig.~\ref{sup_fig:muller_committor_analysis}, we report the results of a standard committor analysis for a set of 500 configurations with $0.45<q<0.55$ sampled with our approach. For each configuration, 100 independent trajectories were run to estimate the corresponding committor value.}

    \subsection{Numerical evaluation of the committor function} \label{sup_sec:numerical_muller}
        One of the advantages of a model system such as the M\"uller-Brown potential is the possibility of solving it numerically to obtain precise reference data.
        In our case, we compute an NN-based committor $q_\theta$ as a function of the Cartesian coordinates ($q_\theta = q_\theta(x,y)$) by applying our method to the ideal dataset obtained from a homogeneous grid (i.e., 200*200 evenly distributed points) in the relevant part of the Cartesian space (i.e.,  -1.4<x<1.1 and -0.25<y<2.0 ).
        At variance with our iterative procedure, in the case of this didactic and ideal scenario, as we know the analytical expression of $U(x,y)$ (see Eq.~\ref{seq:muller_formula}), the weights $w_i$ associated with configuration $i$ in the $L_v$ term of Eq.~\ref{eq:loss_variational}, can be directly computed as the true Boltzmann statistical weight $w_i = e^{-\beta U(x_i,y_i)}$.
        Labeling the data belonging to the metastable states A and B according to the correct basin, we apply the boundary conditions by minimizing the $L_b$ term of Eq.~\ref{eq:loss_boundary} in the same way reported in the main text.
        This way, we can easily optimize the committor $q_\theta$ by minimizing the total loss function of Eq.~\ref{eq:total_loss}.
        For the training, the x and y Cartesian coordinates were used as the input of an NN with architecture [2, 20, 20, 1] that was optimized using the ADAM optimizer. 
        As the result from such a uniformly distributed database is the best result one can get with the same NN architecture, we set the numerical result $q_N$\enrico{, for which we obtained $\mathcal{K}_m=4.18$}, as a reference to test our method (see reference line in Fig.~\ref{fig:muller} in the main text).
        \begin{figure}[h!]
            \centering
            \includegraphics[width=0.5\linewidth]{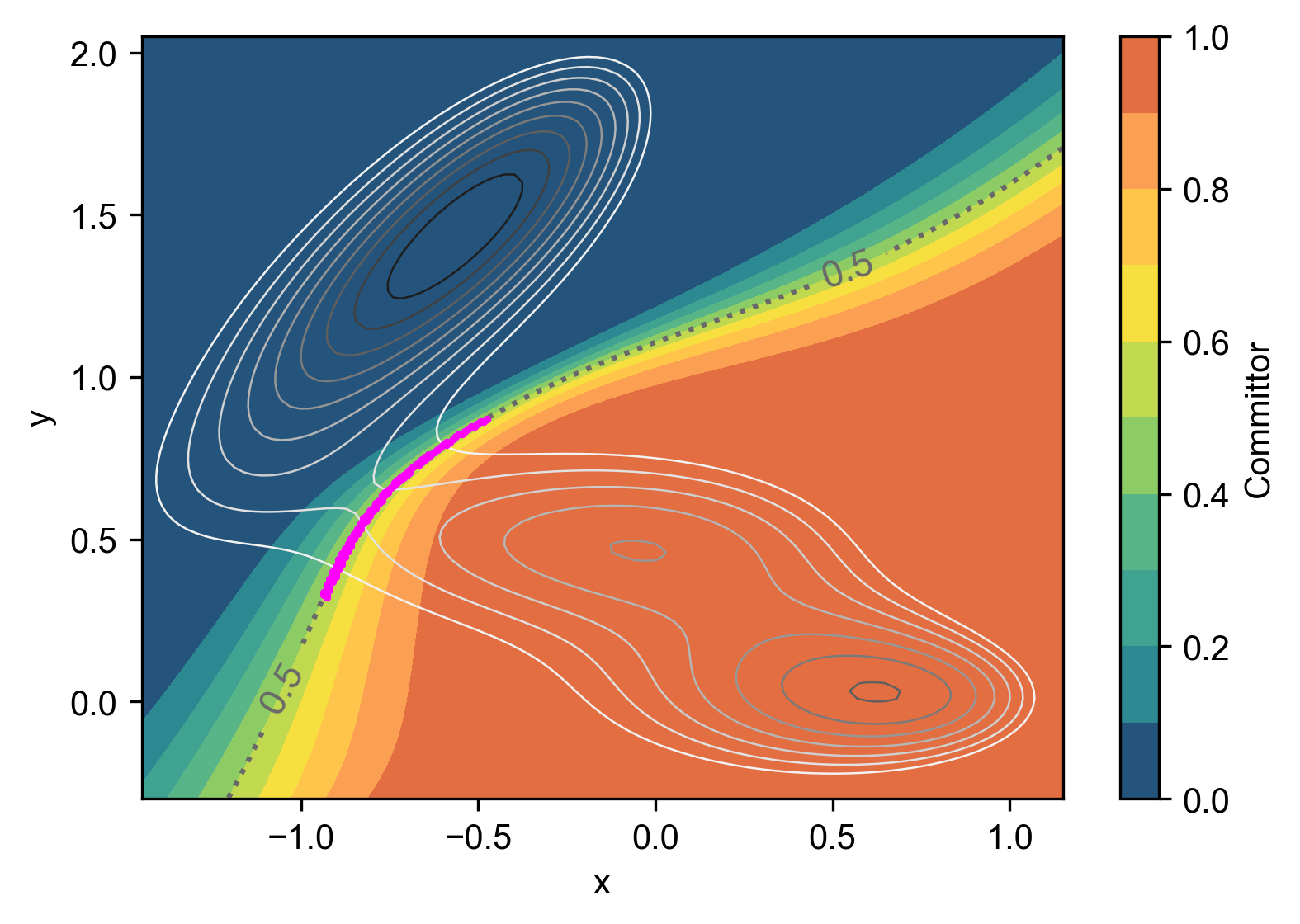}
            \caption{Contour plot of the reference committor function obtained from numerical integration on M\"uller-Brown potential energy surface. \enrico{The full 0.5 isocommittor line is reported as a dotted line, whereas its physically relevant part is highlighted in magenta.}}
            \label{sup_fig:muller_numerical_committor}
        \end{figure}

    \subsection{Additional information on iterative optimization}
        \subsubsection*{Effect of $\,\lambda$ parameter on the bias potential}
            As discussed in the main text, the most useful parameter concerning the sampling under the action of the bias $V_\mathcal{K}(\textbf{x})$ of Eq.~\ref{eq:bias} is the $\lambda$ multiplicative factor that modulates its strength (see Sec.~\ref{subsec:sampling}).            
            An appropriate choice of $\lambda$ allows for balancing $U(\textbf{x})$ and $V_\mathcal{K}(\textbf{x})$ and improving the sampling efficiency. 
            \begin{figure}[h!]
                \centering
                \includegraphics[width=0.9\linewidth]{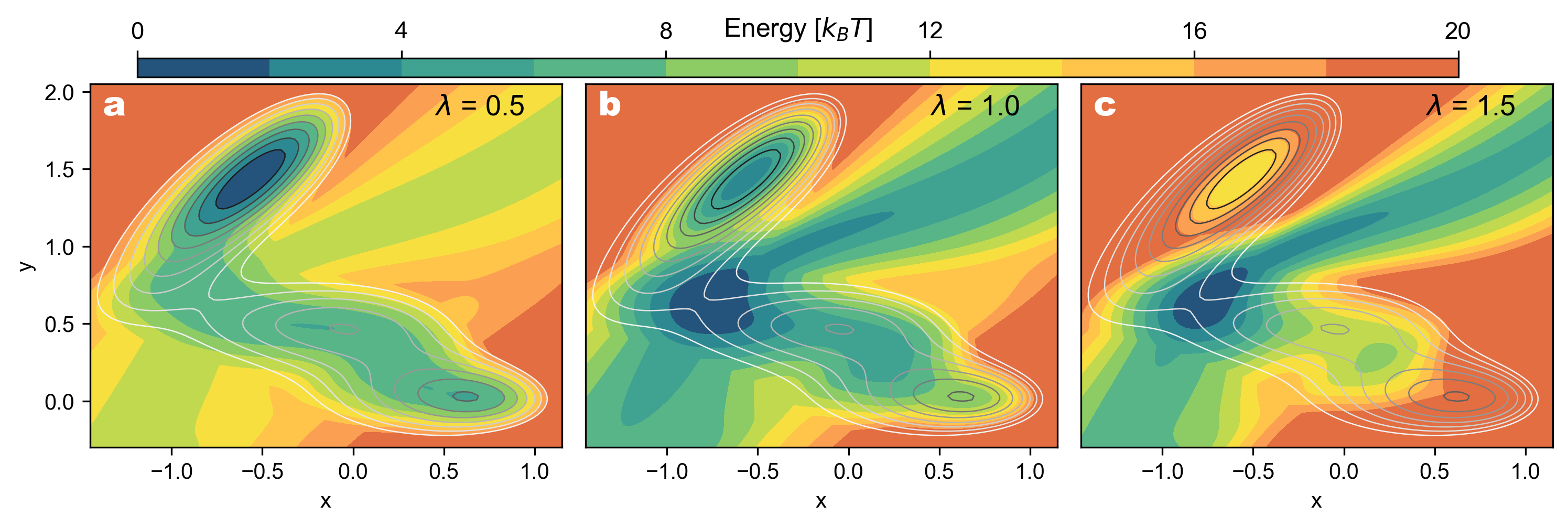}
                \caption{Comparison of the effect of the same $V_K$ bias potential on the M\"uller-Brown potential energy surface as a function of the value of the multiplicative factor $\lambda$, whose value is given in by the top-right black label for each plot.}
                \label{sup_fig:muller_lambda_comparison}
            \end{figure}

            In Fig.~\ref{sup_fig:muller_lambda_comparison}, we report a didactic comparison of the resulting biased energetic landscape under the action of three $V_\mathcal{K}(\textbf{x})$ with $\lambda_a = 0.5$, $\lambda_b = 1.0$, which is the same reported in Fig.~\ref{fig:muller} in the main text, and  $\lambda_c = 1.5$. Even if such a large range of values is not so likely to be used in practice, we report here such values to make the point we want to show more evident.
            
            As the lambda increases, the TS minimum becomes more stable with respect to the real metastable states, which are progressively destabilized.
            This allows for easier sampling of the TS region with shorter escape times from the metastable basins. On the other hand, with a weaker bias, the system may need a long time to escape.
            
            Using a stronger bias can thus be a resource, especially in the earliest iteration of the procedure in which the learned committor model is still to be refined and, as a consequence, the corresponding bias is still rough.         
            However, a stronger bias makes the TS minimum deeper and narrower, somehow limiting the sampling of that region, whereas a milder bias creates a shallower TS minimum, allowing for a broader sampling of configurations from the TS surroundings. 
    
            Even if the choice of $\lambda$ is not so sensitive, as we state in the main text, finding the best trade-off can help speed up sampling.
            In practice, it is good practice to monitor the sampling quality that can be achieved under the action of the $V_\mathcal{K}$ and, based on this feedback, to eventually adjust the value of $\lambda$ to improve the performances. 
            
            In addition, it should also be noted that, as for other enhanced sampling methods based on the addition of an external bias, it is generally better to prefer milder biases to stronger ones when possible, as they can result in unstable simulations or artifacts in the most extreme cases. 

        \subsubsection*{Effect of including $\,\Delta$F information on earlier iteration results}
\begin{figure}[h!]
                \centering
                \includegraphics[width=0.7\linewidth]{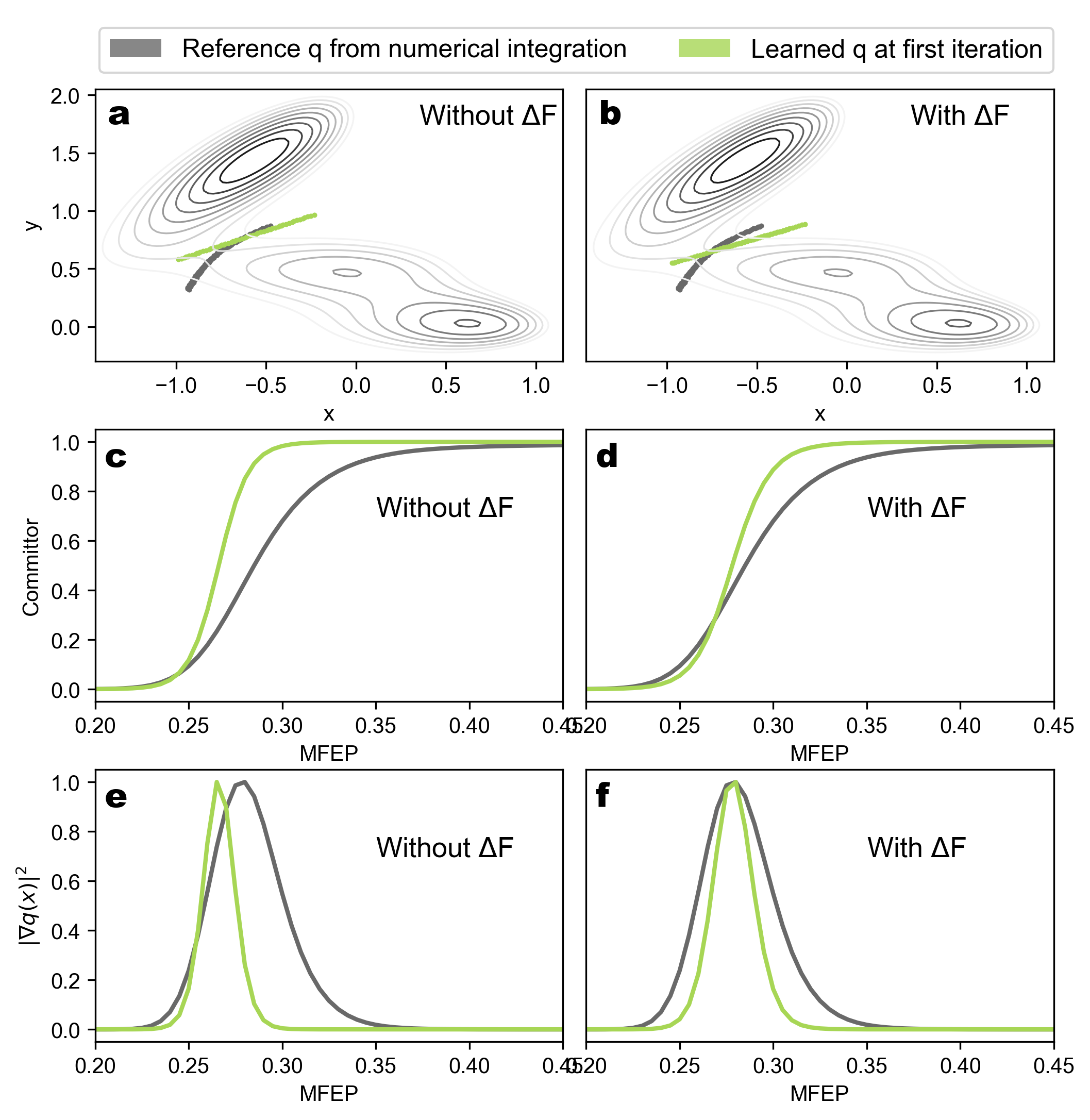}
                \caption{Comparison of the effect of including information about the $\Delta$F between the metastable states on the committor learned at the first iteration for the M\"uller-Brown system. The results are compared with the reference numerical result both in the x and y plane (panels \textbf{a} and \textbf{b}) and as projected along the minimum free energy path (MFEP) (panels \textbf{c} and \textbf{d}).
                In panels \textbf{e} and \textbf{f}, we stress the difference between the two learned models by comparing the value of the term $|\nabla q(\textbf{x})|^2$ along the MFEP section associated to the TS region.}
                \label{sup_fig:muller_deltaf_comparison}
            \end{figure}

            As we discussed in the main text, if an estimate of the free energy difference $\Delta F_{AB}$ is available, the weights of the unbiased data from the first iteration can be corrected to better resemble the true Boltzmann probability.
            Even if not strictly necessary, this allows for speeding up the overall optimization procedure, as it is illustrated in Fig.~\ref{sup_fig:muller_deltaf_comparison}, where we compare the committor functions learned after the first iteration including and not including the $\Delta$F information. 
            Without the additional information, it can be seen that the iscommittor line at $q=0.5$ is far from the TS region and the reference. 
            On the other hand, if the $\Delta$F information is included, the model is already closer to the reference value even from the first iteration.
            
            The effect of $\Delta F$ can also be seen in the final converged Kolmogorov distribution in Fig.~\ref{sup_fig:muller_deltaf_comparison}. Although we can already get a good isocommittor line at $q=0.5$ in 3 iterations, as we reported in the main text, the whole Kolmogorov distribution is not perfect yet if compared with the one obtained from numerical integration (see Sec.~\ref{sup_sec:numerical_muller}). 
            To match such an ideal result, the iterative procedure needs 14 cycles without $\Delta F_{AB}$ and only 6 cycles with $\Delta F_{AB}$.

            However, as already stated in the main text, the effect of such additional information has relative importance in the overall results in both cases, as it mostly affects minor details.
            \begin{figure}[h!]
                \centering
                \includegraphics[width=0.75\linewidth]{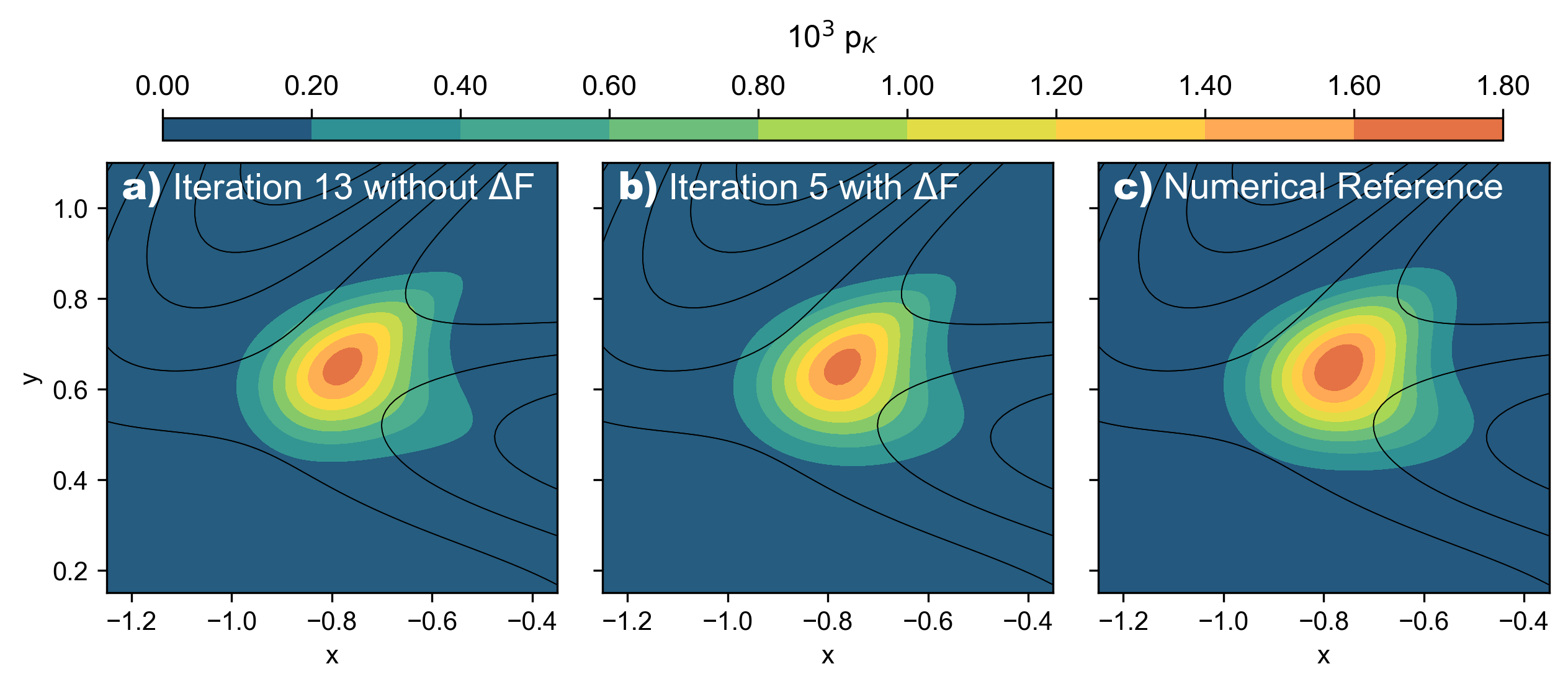}
                \caption{Comparison of the number of iterations needed to converge to the reference Kolmogorov distribution $p_\mathcal{K}$ including (6 iterations) and not including (14 iterations) information about the $\Delta F$. It should be noted that the convergence of the overall Kolmogorov distribution is slower than the iscommittor line $q=0.5$ as it requires more information. \enrico{The $\mathcal{K}_m$ values obtained with the models reported in panels \textbf{a} and \textbf{b} are 4.31 and 4.20, respectively.}}
                \label{sup_fig:muller_kolmogorov_comparison}
            \end{figure}
            
\clearpage
\section{Alanine Dipeptide - Additional information}
    \subsection{Computational details}
        \subsubsection*{Simulations details}
            The alanine dipeptide (Ace-Ala-Nme) simulations in vacuum have been carried out using the GROMACS-2021.5~\cite{abraham2015gromacs} MD engine patched with PLUMED~\cite{tribello2014plumed, plumed2019promoting} and the Amber99-SB~\cite{amber2013} force field with a 2 fs timestep. The Langevin dynamics is sampled with damping coefficient $\gamma_i= \frac {m_i}{\tau-t}$ with $\tau-t = 0.05$ ps and target temperature 300 K.

    \subsubsection*{Committor model training details}
        For modeling the NN-based committor function $q_\theta(\textbf{d}(\textbf{x}))$ for alanine dipeptide, we tested a set of possible descriptors $d(\textbf{x})$, which are reported in Table~\ref{sup_tab:alanine_descriptors}.
        In all the cases, we kept the NN architecture similar for consistency, just changing the size of the input layer to match the number $N_\textbf{d}$ of descriptors used and keeping the same hidden layers, i.e., [$N_\textbf{d}$, 32, 32, 1] nodes/layer.         
        For the optimization, we used the ADAM optimizer with an initial learning rate of $10^{-3}$ modulated by an exponential decay with multiplicative factor $\gamma=0.99999$. 
        The training was performed for $\sim$30000 epochs. The $\alpha$ hyperparameter in Eq.~\ref{eq:total_loss} was set to 10.

        In Table~\ref{sup_tab:alanine_descriptors}, we compare the performances of the different descriptors in terms of their capability of minimizing the variational loss term. To standardize this result, we performed the optimization of $K_{m}$ on the same configurations dataset for all the descriptors. The reported errors are computed as the standard deviations on three trained models with different random initializations of the weights.
        
        \begin {table}[h!]
            \caption {Comparison of the performances of different descriptor sets for alanine dipeptide. The descriptor sets discussed in the main text are marked with an asterisk.} \label{sup_tab:alanine_descriptors} 
            \begin{center}
            \begin{tabular}{ |c|c| } 
                 \hline
                Descriptors \textbf{x}         &   $K_{m}$    \\
                \hline
                $ \phi$                        & $8.8 \pm 0.2$  \\
                $\phi, \psi^*$                 & $7.6 \pm 0.1$  \\ 
                $\phi,\omega$                  & $8.2 \pm 0.4$  \\
                $\phi,\psi, \omega$            & $7.1 \pm 0.2$  \\
                $\phi,\theta^*$                & $3.4 \pm 0.1$  \\ 
                $\phi, \psi, \theta$           & $2.7 \pm 0.1$  \\
                $\phi,\theta, \omega$          & $3.0 \pm 0.1$  \\
                $\phi,\psi,\theta, \omega$     & $2.6 \pm 0.1$  \\
                45 distances$^*$               & $1.1 \pm 0.1$  \\
                $O_{proj}$                     & $1.2 \pm 0.1$  \\ [1ex]
            \hline
            \end{tabular}
            \end{center}
        \end {table}

        The number of iterations, the corresponding dataset size, and the $\lambda$ used in the biased simulations are summarized in Table~\ref{sup_tab:alanine_iterations} alongside the lowest value obtained for the variational loss term $L_v$, which provides a quality and convergence measure\enrico{, the simulation time $t_s$ and the output sampling time $t_o$}.
        \begin {table}[h!]
            \caption {Summary of iterative procedure for Alanine (distance model)} \label{sup_tab:alanine_iterations}
            \begin{center}
            
            \begin{tabular}{ |c|c|c|c|c|c| } 
             \hline
             Iteration & Dataset size & $K_{m}$ & $\lambda$ & \enrico{$t_s$ [ns]} & \enrico{$t_o$ [ps]} \\ 
             \hline
                0   & 20000  &  43.7 & -   & \enrico{2*4} & \enrico{0.4} \\
                1   & 40000  &  4.9  & 0.8 & \enrico{2*4} & \enrico{0.4} \\
                2   & 60000  &  1.3  & 0.8 & \enrico{2*4} & \enrico{0.4} \\
                3   & 80000  &  1.2  & 0.8 & \enrico{2*4} & \enrico{0.4} \\
                4   & 100000 &  1.1  & 0.8 & \enrico{2*4} & \enrico{0.4} \\
                5   & 120000 &  1.1  & 0.8 & \enrico{2*4} & \enrico{0.4} \\ [1ex] 
             \hline
            \end{tabular}
            \end{center}
        \end {table}

\begin{figure}[h!]
            \centering
            \includegraphics[width=1\linewidth]{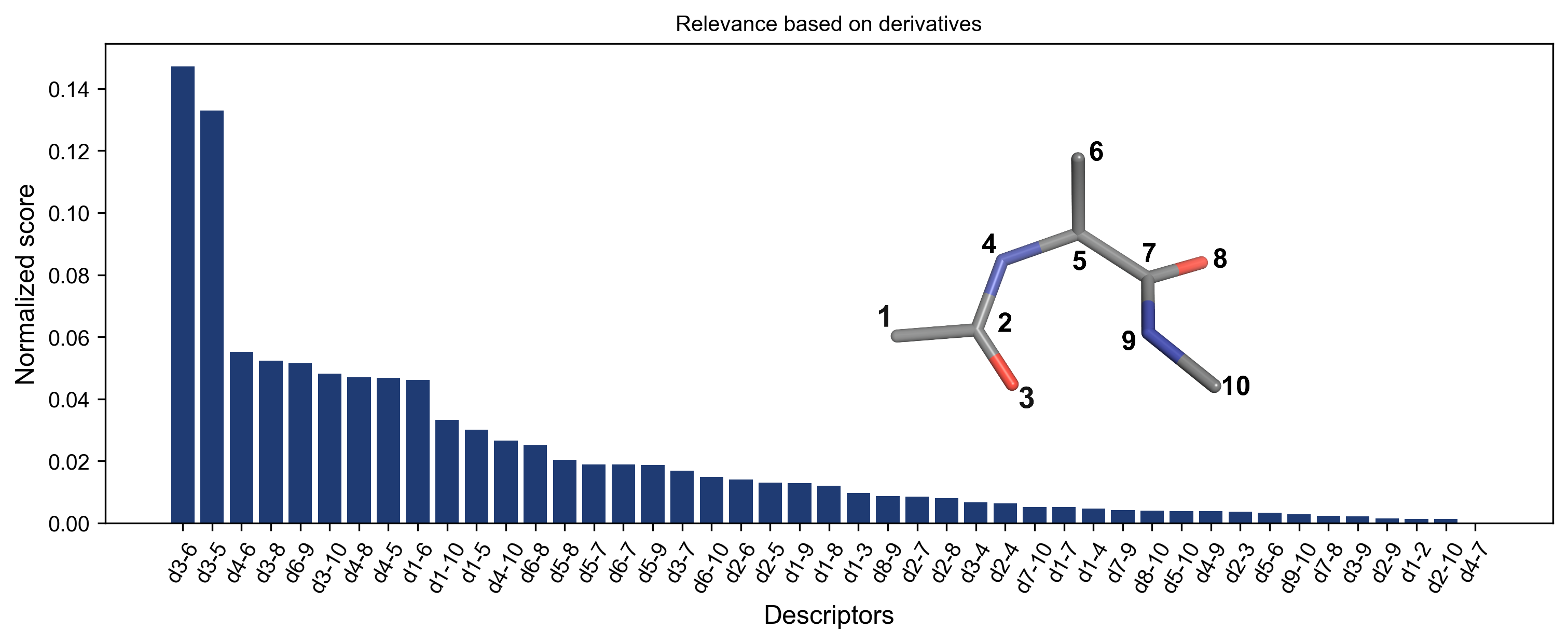}    
            \caption{Descriptors ranking (see Sec.~\ref{sup_sec:feature_ranking}) for the committor model of alanine dipeptide trained using the 45 distances between heavy atoms as inputs. The descriptors are named based on the labels on the molecule provided in the inset.}
            \label{sup_fig:alanine_rank}
        \end{figure}

        In Fig.~\ref{sup_fig:alanine_rank}, we report the ranking of the input for the committor model based on the 45 distances input descriptors set computed as described in Sec.~\ref{sup_sec:feature_ranking}. This shows that the most relevant descriptors in our model are the $d_{36}$ and $d_{35}$, which are the $d_\beta$ and $d_\alpha$ reported in the main text, respectively. 
        
\begin{figure}[h!]
            \centering
            \includegraphics[width=0.4\linewidth]{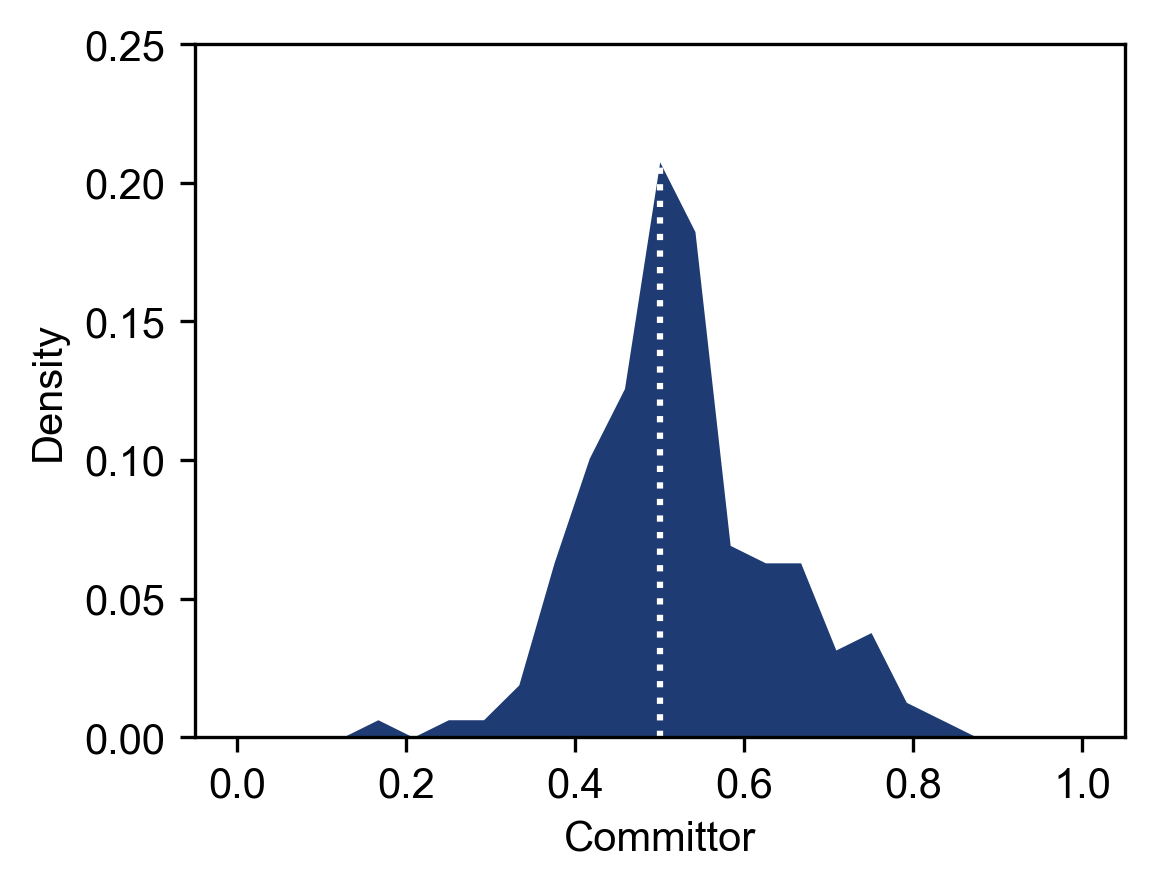}    
            \caption{\enrico{Normalized distribution of the results of the committor analysis for a set of 100 alanine configurations with $0.45<q<0.55$ sampled with our approach.}}
            \label{sup_fig:alanine_committor_analysis}
        \end{figure}

        \enrico{In Fig.~\ref{sup_fig:alanine_committor_analysis}, we report the results of a standard committor analysis for a set of 100 configurations with $0.45<q<0.55$ sampled with our approach. For each configuration, 100 independent trajectories were run to estimate the corresponding committor value.}

        \subsection{Short note on the transition state ensemble}
\begin{figure}[h!]
            \centering
            \includegraphics[width=0.8\linewidth]{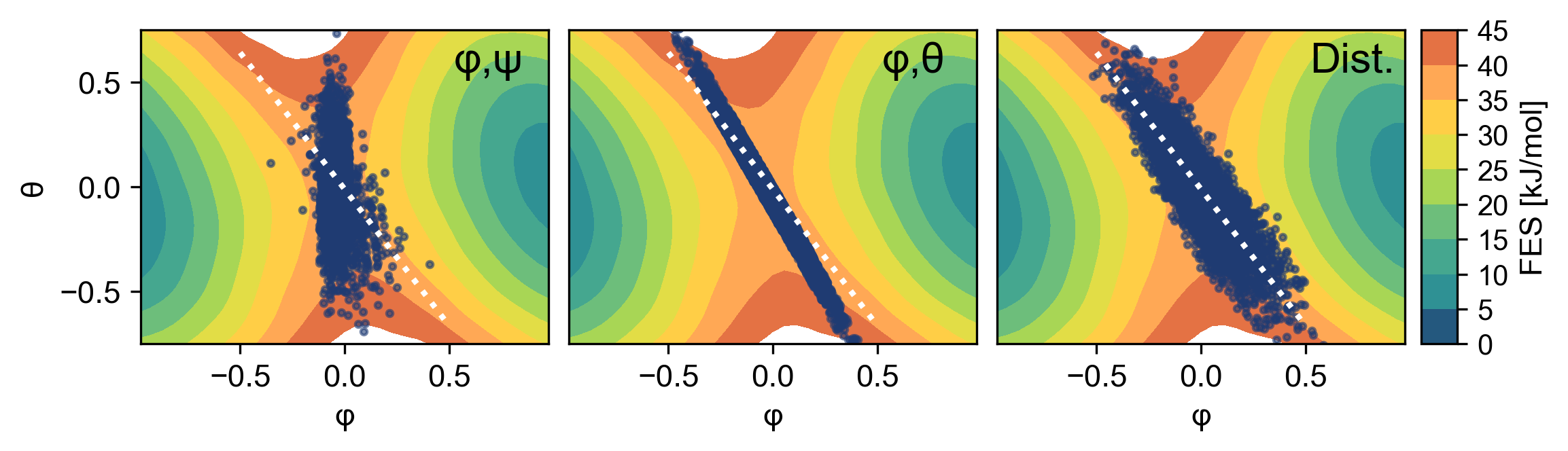}
            \caption{Scatter plot of the points for which 0.4<q<0.6 according to three committor models with different inputs $\phi \psi$ angles, $\phi \theta$ angles, and the set of 45 distances. The underlying FES is depicted by the colormap, the white dashed lines report the reference linear relation for the TS between $\phi$ and $\theta$. The reported region of the $\phi\theta$ space is the same as panel \textbf{c} of Fig.~\ref{fig:alanine} in the main text to which this figure should be compared. }
            \label{sup_fig:alanine_TS_from_q}
        \end{figure}

\begin{figure}[h!]
            \centering
            \includegraphics[width=0.8\linewidth]{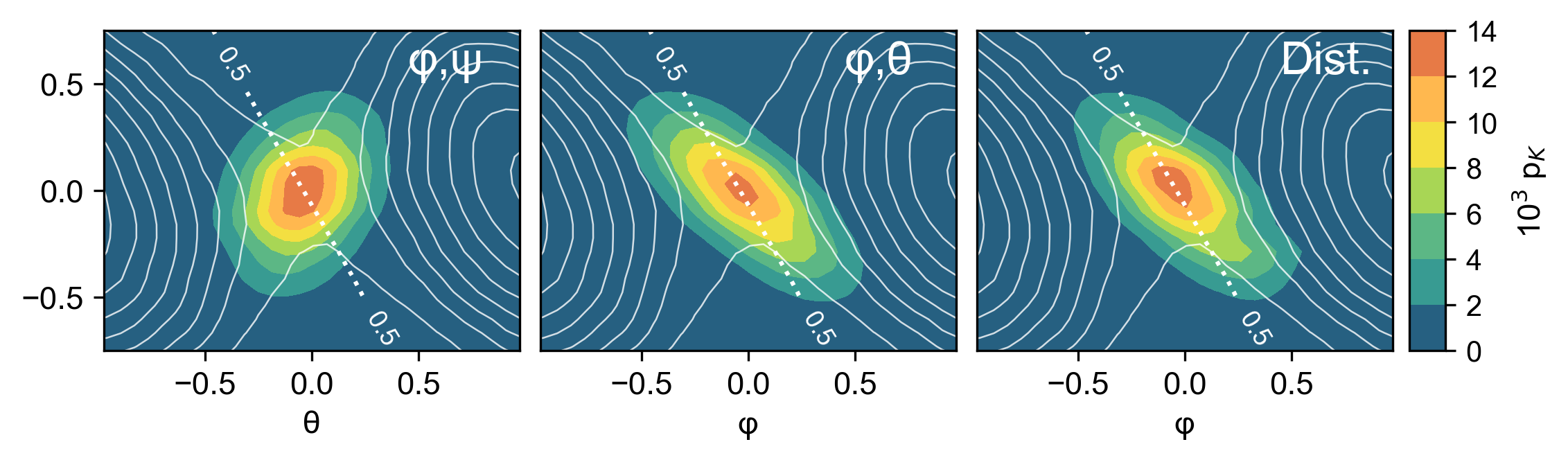}
            \caption{\enrico{Contour plot of $p_K$ according to three committor models with different inputs $\phi \psi$ angles, $\phi \theta$ angles, and the set of 45 distances. The underlying FES is depicted by the white isolines, the white dashed lines report the $\simeq$0.5 isoline of the committor learned as a function of $\phi$ and $\theta$. The reported region of the $\phi\theta$ space is the same as panel \textbf{c} of Fig.~\ref{fig:alanine} in the main text to which this figure should be compared. }}
            \label{sup_fig:alanine_pk_vs_q05}
        \end{figure}

        In the main text, in panel \textbf{c} of Fig.~\ref{fig:alanine}, we report the TSE on the basis of our novel Kolmogorov distribution $p_\mathcal{K}(\textbf{x})$ according to different models with different inputs.
        Here in Fig.~\ref{sup_fig:alanine_TS_from_q}, for comparison, we follow the conventional approach of selecting TS-related configuration as belonging to the surroundings of the isosurface $q=0.5$. We plot the results on the same region of the $\phi\theta$ space as panel \textbf{c} of Fig.~\ref{fig:alanine}.
        \enrico{Our definition of the TSE based on $p_K$, however, shouldn't be seen as in contrast to the conventional definition based on the $q\simeq0.5$ criterion, but rather as an extension of such a concept. Indeed, as we show in Fig.~\ref{sup_fig:alanine_pk_vs_q05}, the maximum of $p_K$ is found in correspondence with the lowest energy point on the $q\simeq0.5$ isoline.}        

\begin{figure}[h!]
            \centering
            \includegraphics[width=0.8\linewidth]{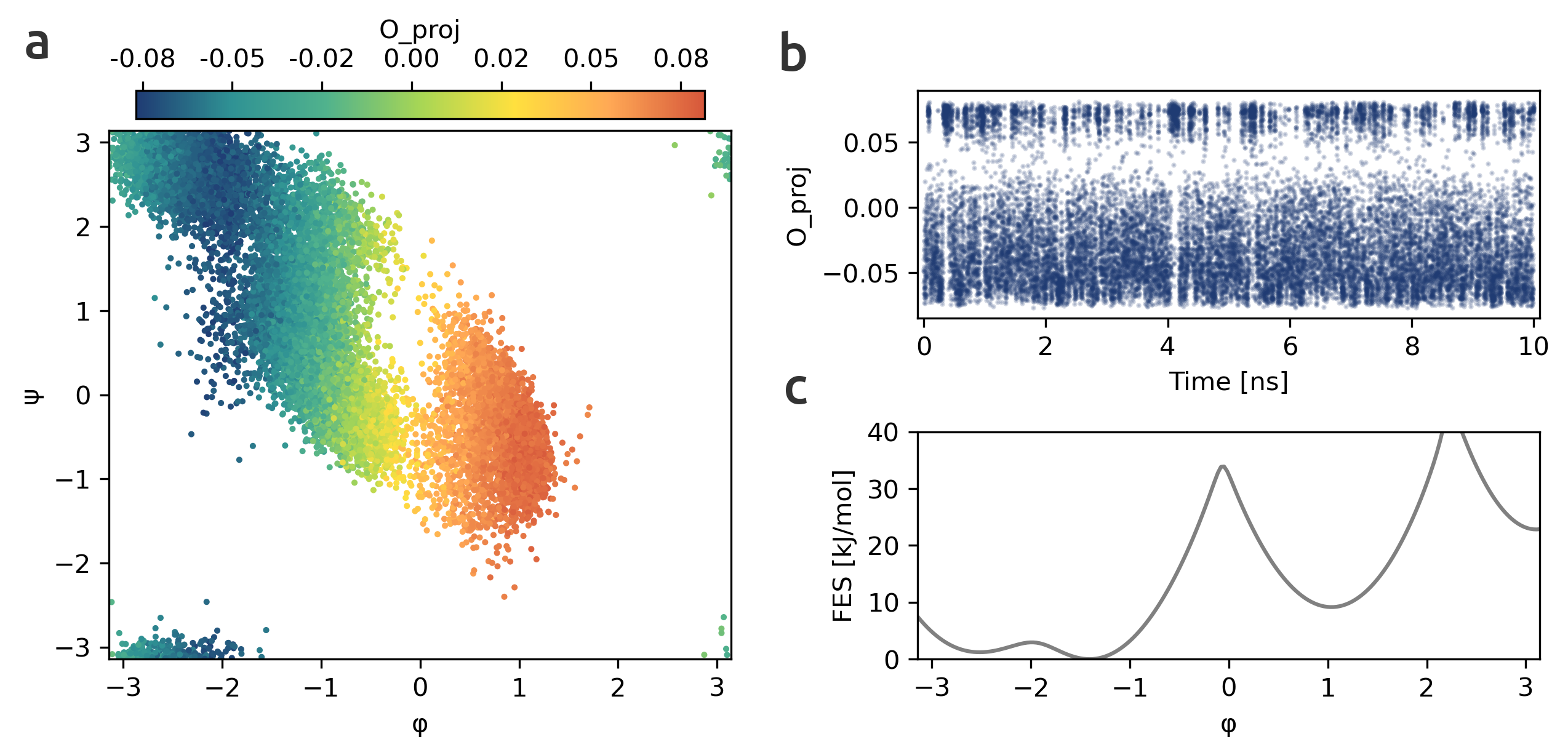}
            \caption{\enrico{Results of OPES sampling of alanine using the O\_projection CV described in the main text. \textbf{a)} Scatter plot of the sampled points in the $\phi$,$\psi$ plane colored according to the value of the O\_projection value \textbf{b)} Time series of the O\_projection variable \textbf{c)} Free energy surface (FES) computed along the reference $\phi$ torsional angle.}}
            \label{sup_fig:alanine_o_projection}
        \end{figure}

\enrico{\subsection{O projection as a CV for enhanced sampling simulations of alanine conformational equilibrium}
    To quickly check how representative the projection of the position of the O atom on the NCC$_\beta$ plane could be for alanine conformational equilibrium, we performed a simple biased simulation along such a CV, using the On-the-fly Probability Enhanced Sampling~\cite{invernizzi2020rethinking} (OPES) method, which is a recent development of Metadynamics~\cite{laio2002escaping}.
    From the scatter plot and the time series in Fig.~\ref{sup_fig:alanine_o_projection}, panels a and b, respectively, it is evident that it is an effective CV for the system, thus promoting many transitions between the two basins.
    The OPES parameters in PLUMED~\cite{plumed2019promoting} for this simulation were: \texttt{BARRIER=25}, \texttt{PACE=500}, \texttt{SIGMA=0.002}. The O\_projection CV was implemented in PyTorch and deployed to PLUMED using the \texttt{PYTORCH\_MODEL} interface~\cite{bonati2023mlcolvar} taking the positions of the involved atoms (ONCC$_\beta$) as inputs.}
    \label{sup_sec:o_projection}

\clearpage
\section{DASA reaction - Additional information}
    \subsection{Computational details}
        \subsubsection*{Simulations details}
            The DASA reaction simulations have been carried out using the CP2K-8.1~\cite{cp2k2020} software package at PM6 semi-empirical level~\cite{stewart2007optimization}. 
            The integration step was $0.5$ fs, and we used the velocity rescaling thermostat~\cite{bussi2007velocity} set at 300K with a time constant of 100 fs.

        \subsubsection*{Committor model training details}
            To model the committor function $q_\theta(\textbf{x})$ at each iteration, we used the 45 distances between the 9 heavy atoms involved in the reaction plus the H involved in the proton transfer (see labeled atoms in Fig.~\ref{fig:dasa} in the main text) as inputs of a neural network (NN) with architecture [45, 32, 32, 1] nodes/layer.
            For the optimization, we used the ADAM optimizer with an initial learning rate of $10^{-3}$ modulated by an exponential decay with multiplicative factor $\gamma=0.99999$. 
            The training was performed for $\sim$20000 epochs.  The $\alpha$ hyperparameter in Eq.~\ref{eq:total_loss} was set to 10.
            
            The number of iterations, the corresponding dataset size, and the $\lambda$ used in the biased simulations are summarized in Table~\ref{sup_tab:dasa_iteration} alongside the lowest value obtained for the functional $K_{m}$ , which provides a quality and convergence measure\enrico{, the simulation time $t_s$ and the output sampling time $t_o$}.
            The reported errors are computed as the standard deviations on three trained models with different random initializations of the weights.

            \begin {table}[h!]
                \caption {Summary of the iterative procedure for DASA reaction.} \label{sup_tab:dasa_iteration}
                \begin{center}
                \begin{tabular}{ |c|c|c|c|c|c| } 
                 \hline
                 Iteration & Dataset size & $K_{m}$  & $\lambda$ & \enrico{$t_s$ [ps]} & \enrico{$t_o$ [fs]} \\ 
                 \hline
                    0   & 10000 &  166311 & -       & \enrico{2*100} & \enrico{10}\\
                    1   & 50000 &  267   & 2.4-3.2  & \enrico{2*100-2*100} & \enrico{10} \\
                    2   & 52000 &   35295 & 1  & \enrico{2*10} & \enrico{10}\\
                    3   & 54000 &  5.29 &   1  & \enrico{2*10} & \enrico{10}\\
                    4   & 56000 &  6.08 & 1.2  & \enrico{2*10} & \enrico{10}\\
                    5   & 58000 &  8.71 & 1.2  & \enrico{2*10} & \enrico{10}\\   
                    6   & 60000 &  2.28 & 1.2  & \enrico{2*10} & \enrico{10}\\
                    7   & 62000 &  6.83 & 1.2  & \enrico{2*10} & \enrico{10}\\
                    8   & 64000 &  2.26 & 1.2  & \enrico{2*10} & \enrico{10}\\  
                    9   & 66000 &  2.14 & 1.2  & \enrico{2*10} & \enrico{10}\\
                    10  & 68000 &  1.49 & 1.2  & \enrico{2*10} & \enrico{10}\\  
                    11  & 76000 &  1.44 & 1.2  & \enrico{4*10} & \enrico{10}\\  
                 \hline
                \end{tabular}
                \end{center}
            \end {table}

\begin{figure}[h!]
                \centering
                \includegraphics[width=1\linewidth]{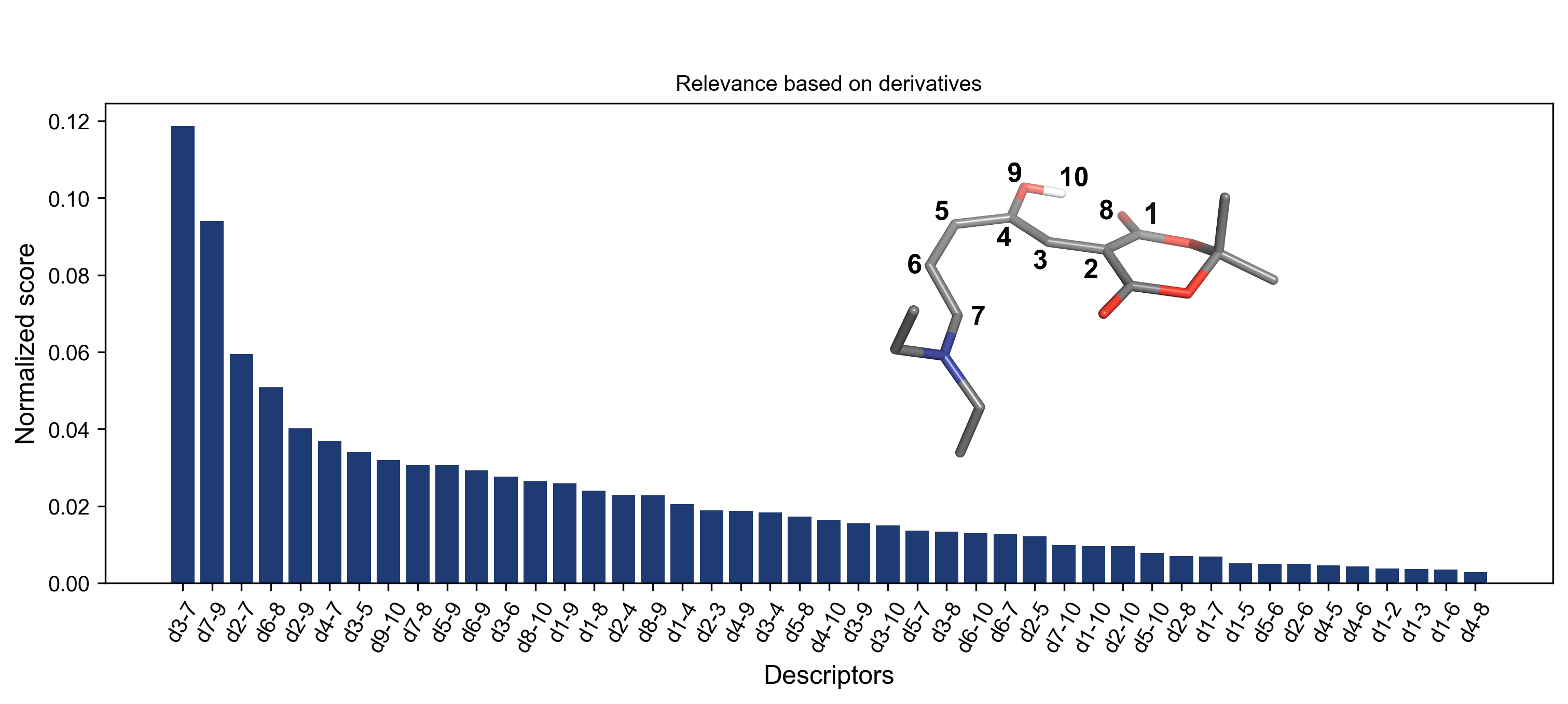}
                \caption{Descriptors ranking (see Sec.~\ref{sup_sec:feature_ranking}) for the committor model of DASA reaction trained using the 45 distances between heavy atoms plus the proton transfer H atom as inputs. The descriptors are named based on the labels on the molecule provided in the inset.}
                \label{sup_fig:dasa_rank}
            \end{figure}

            In Fig.~\ref{sup_fig:dasa_rank}, we report the ranking of the input for the committor model based on the 45 distances input set, computed as described in Sec.~\ref{sup_sec:feature_ranking}. 
            A discussion of such results can be found in the main text.

\begin{figure}[h!]
                    \centering
                    \includegraphics[width=0.4\linewidth]{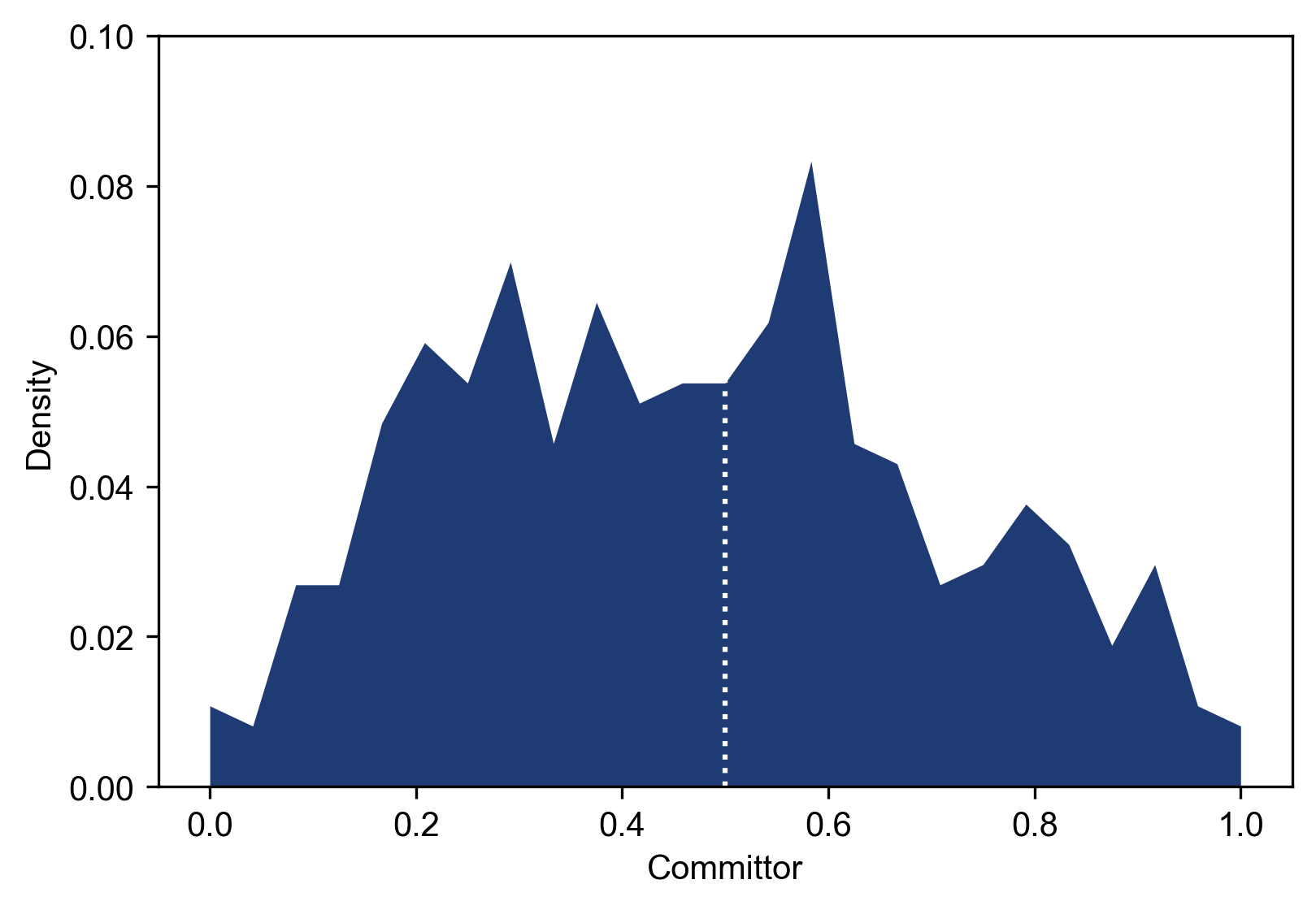}    
                    \caption{\enrico{Normalized distribution of the results of the committor analysis for a set of 300 DASA reaction configurations with $0.45<q<0.55$ sampled with our approach.}}
                    \label{sup_fig:dasa_committor_analysis}
                \end{figure}

            \enrico{In Fig.~\ref{sup_fig:dasa_committor_analysis}, we report the results of a standard committor analysis for a set of 300 configurations with $0.45<q<0.55$ sampled with our approach. For each configuration, 50 independent trajectories were run to estimate the corresponding committor value.}

\begin{figure}
                \centering
                \begin{minipage}{0.48\linewidth}
                    \includegraphics[width=0.6\linewidth]{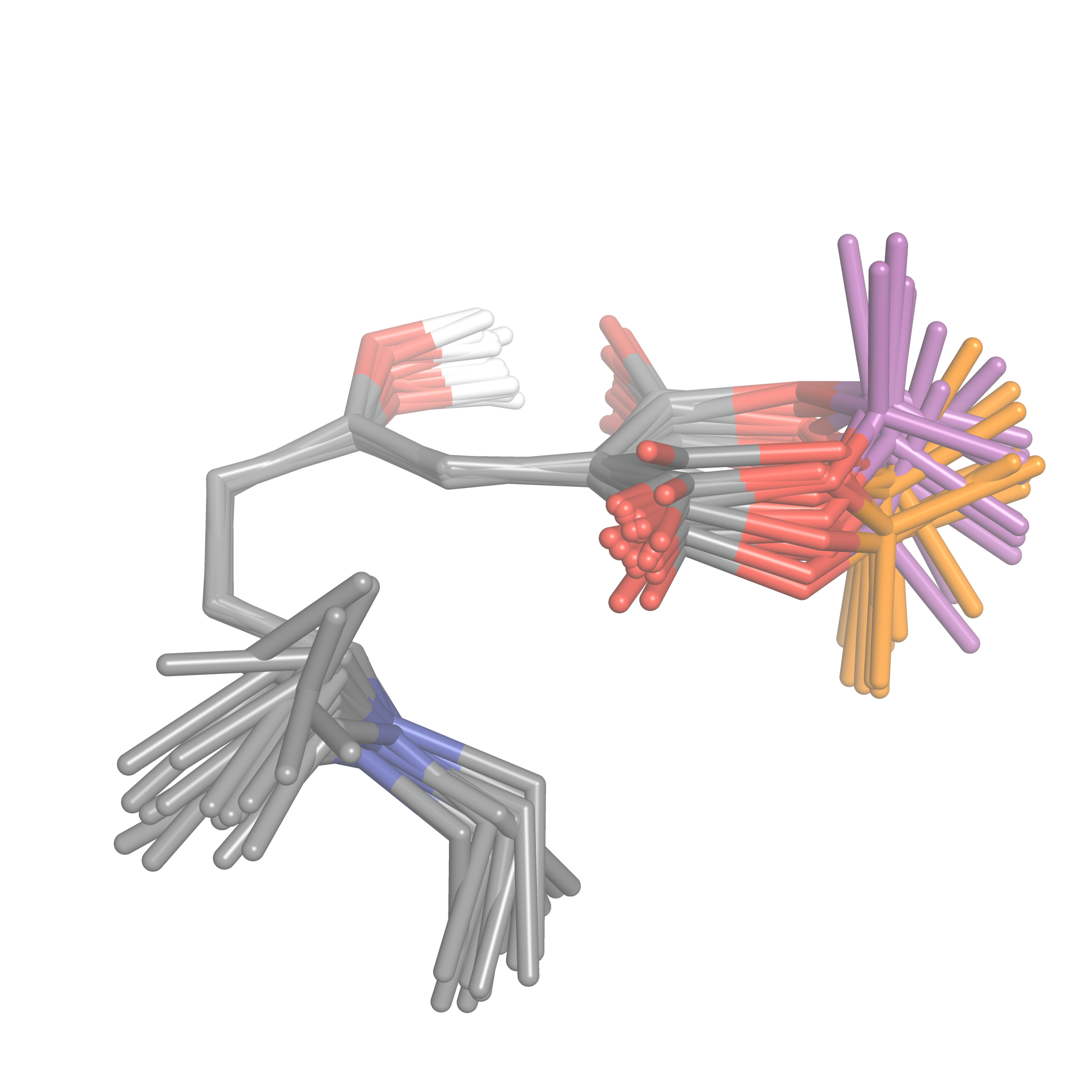}
                \end{minipage}
                \begin{minipage}{0.5\linewidth}
                    \includegraphics[width=0.8\linewidth]{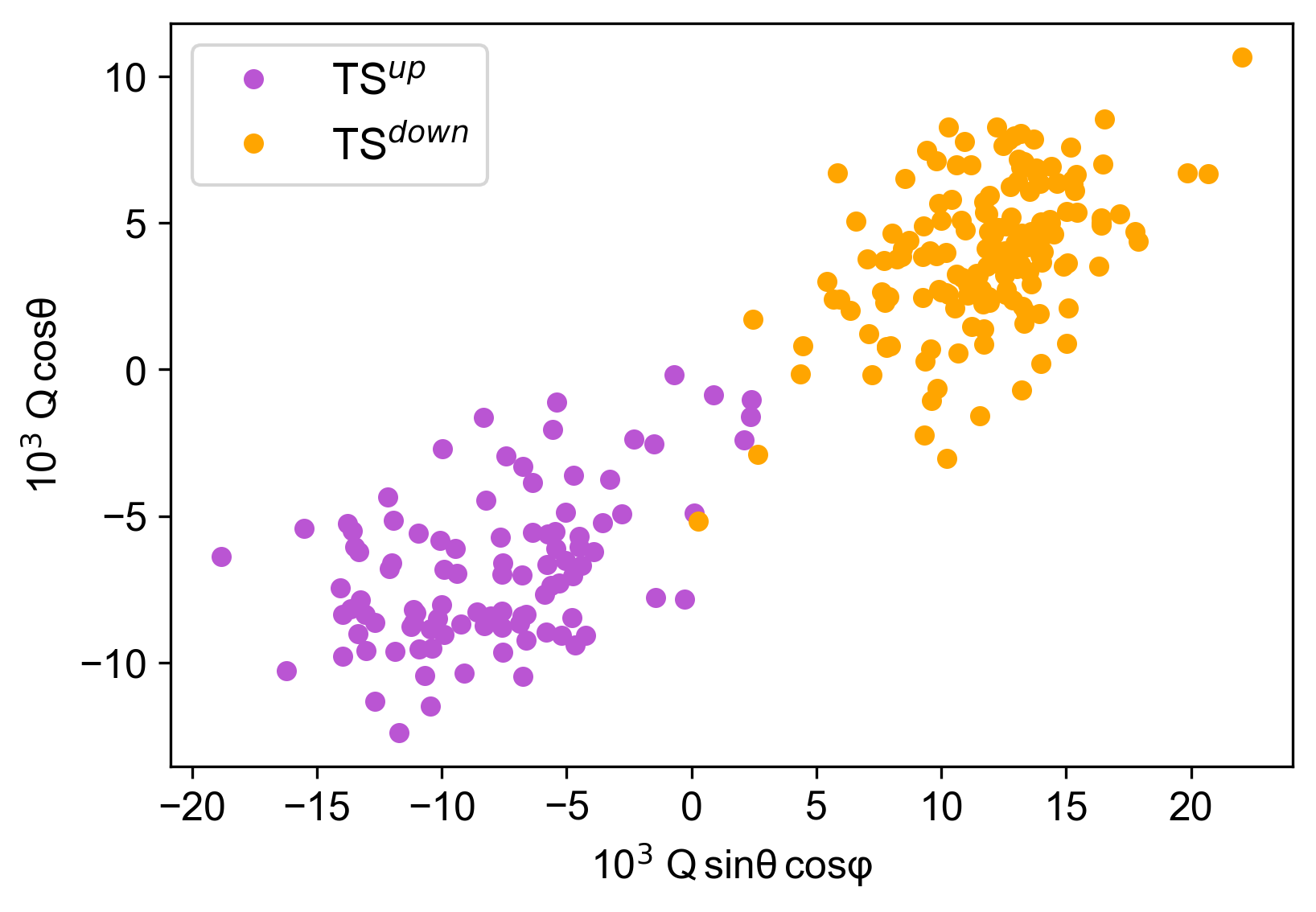}
                \end{minipage}
                \caption{Superimposition of TSE configurations for the DASA reaction and projection of two different TS clusters obtained with k-medoids method on puckering coordinates~\cite{cremer1975general} as described in the main text.}
                \label{sup_fig:dasa_TSall}
            \end{figure}

            In Fig.~\ref{sup_fig:dasa_TSall}, we report a superimposition of the collected TSE configurations for the DASA reaction and the projection of the two clusters identified via the k-medoid analysis on the puckering coordinates.
            A discussion of such results can be found in the main text.

\begin{figure}
                \centering
                \includegraphics[width=0.4\linewidth]{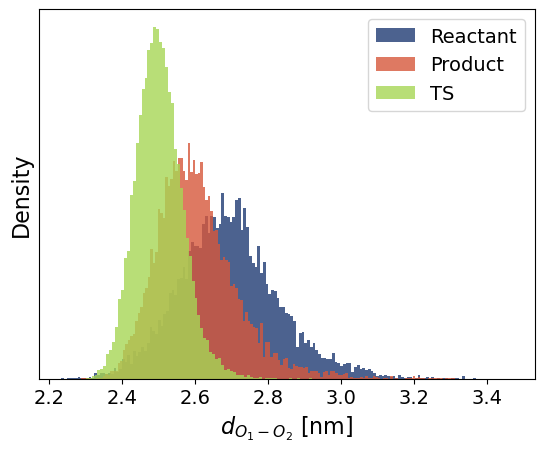}
                \caption{Normalized distribution of the distance between the O atoms involved in the proton transfer (O$_1$O$_2$) in DASA reaction sampled in the reactant state, product state, and the transition state (TS).}
                \label{sup_fig:dasa_distribution_dOO}
            \end{figure}

            In Fig.~\ref{sup_fig:dasa_distribution_dOO}, we report the distribution of O$_1$O$_2$ distance, which involves the O atoms that take part in the proton transfer. In the TSE configurations, the O$_1$O$_2$ distance is reduced due to the conformational change that is needed to facilitate the proton transfer.

\clearpage
\section{Chignolin - Additional information}
    \subsection{Computational details}
        \subsubsection*{Simulations details}
            For the study of folding and unfolding of chignolin (CLN025 peptide sequence Tyr-Tyr-Asp-Pro-Glu-Thr-Gly-Thr-Trp-Tyr) in explicit solvent, we performed our simulations using GROMACS v2021.5~\cite{abraham2015gromacs} patched with PLUMED~\cite{tribello2014plumed, plumed2019promoting}, the CHARMM22$^*$~\cite{piana2011robust} force field, and the solvent has been modeled by the CHARMM TIP3P~\cite{mackerell1998tip} force field, sharing the same setup used for long unbiased simulations on this system~\cite{lindorff2011fast} to have a direct comparison with those results. 
            For the same reason, we kept the simulation condition consistent with that work.
            All simulations were performed with an integration time step of 2 fs and sampling NVT ensemble at 340K.
            Asp, Glu residues, as well as the N- and C-terminal amino acids are simulated in their charged states. The simulation box contains 1,907 water molecules, together with two sodium ions that neutralize the system. The linear constraint solver algorithm is applied to every bond involving H atoms, and electrostatic interactions are computed via the particle mesh Ewald scheme, with a cutoff of 1 nm for all nonbonded interactions.

        \subsubsection*{Committor model training details}
            To model the committor function $q_\theta(\textbf{x})$ at each iteration, we used the 45 distances between the distances between the 10 $\alpha$-carbons of the protein as inputs of a neural network (NN) with architecture  [45, 32, 32, 1] nodes/layer.
            For the optimization, we used the ADAM optimizer with an initial learning rate of $10^{-3}$ modulated by an exponential decay with multiplicative factor $\gamma=0.9999$. 
            The training was performed for $\sim$30000 epochs. The $\alpha$ hyperparameter in Eq.~\ref{eq:total_loss} was set to 10.
            
            The number of iterations, the corresponding dataset size, and the $\lambda$ used in the biased simulations are summarized in Table~\ref{sup_tab:dasa_iteration} alongside the lowest value obtained for the functional $K_m$, which provides a quality and convergence measure, the simulation time $t_s$ and the output sampling time $t_o$.
            The reported errors are computed as the standard deviations on three trained models with different random initializations of the weights.
                    
            \begin {table}[h!]
                \caption {Summary of the iterative procedure for chignolin.} \label{sup_tab:chignolin_iteration}
                \begin{center}
                \begin{tabular}{ |c|c|c|c|c|c| } 
                 \hline
                 Iteration & Dataset size & $K_m$ & $\lambda$ & \enrico{$t_s$ [ns]} & \enrico{$t_o$ [ps]} \\ 
                 \hline
                    0   & 16000 &  2.52 & -   & \enrico{2*40} & \enrico{5}\\
                    1   & 32000 &  1.09 &  0.5-0.72  & \enrico{2*40} & \enrico{5}\\
                    2   & 48000 &  1.03 &  0.5-0.72  & \enrico{2*40} & \enrico{5}\\
                    3   & 64000 &  0.73 &  0.5-0.72  & \enrico{2*40} & \enrico{5}\\
                    4   & 80000 &  0.68 &  0.5-0.72  & \enrico{2*40} & \enrico{5}\\
                    5   & 96000 &  0.70 &  0.5-0.72  & \enrico{2*40} & \enrico{5}\\   
                    6   & 112000 &  1.01 & 0.5-0.72  & \enrico{2*40} & \enrico{5}\\
                    7   & 128000 &  0.72 & 0.5-0.72  & \enrico{2*40} & \enrico{5}\\
                    8   & 144000 &  0.90 & 0.5-0.72  & \enrico{2*40} & \enrico{5}\\  
                    9   & 160000 &  0.78 & 0.5-0.72  & \enrico{2*40} & \enrico{5}\\
                    10   & 176000 & 0.64 & 0.5-0.72  & \enrico{2*40} & \enrico{5}\\  
                 \hline
                \end{tabular}
                \end{center}
            \end {table}

\begin{figure}
                \centering
                \includegraphics[width=1\linewidth]{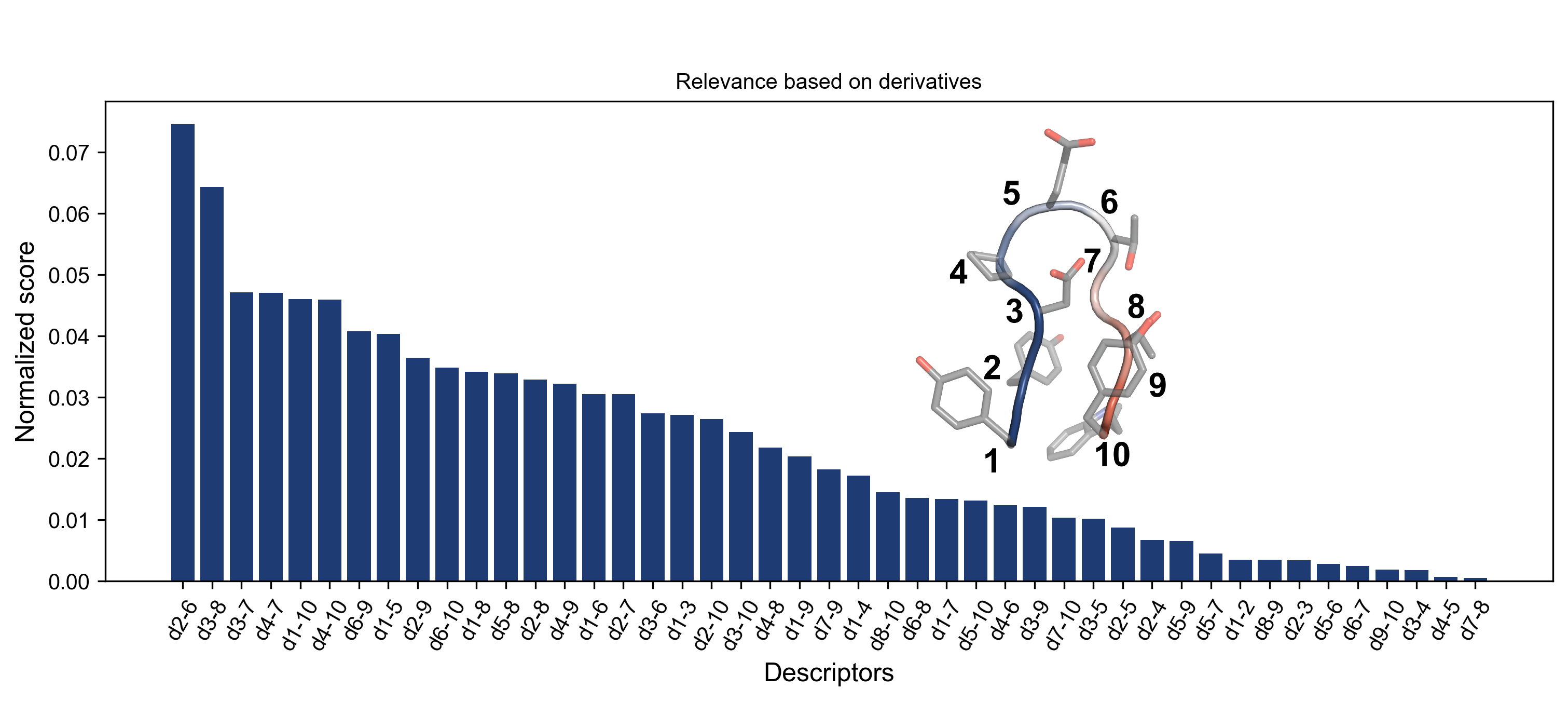}
                \caption{Descriptors ranking (see Sec.~\ref{sup_sec:feature_ranking}) for the committor model of chignolin folding trained using the 45 distances between the $\alpha$-carbons of the protein as inputs.
                The descriptors are named based on the labels on the molecule provided in the inset.}
                \label{sup_fig:chignolin_rank}
            \end{figure}

            In Fig.~\ref{sup_fig:chignolin_rank}, we report the ranking of the input for the committor model based on the 45 distances input set, computed as described in Sec.~\ref{sup_sec:feature_ranking}. 
            A discussion of such results can be found in the main text.
            
\begin{figure}
                \centering
                \includegraphics[width=0.4\linewidth]{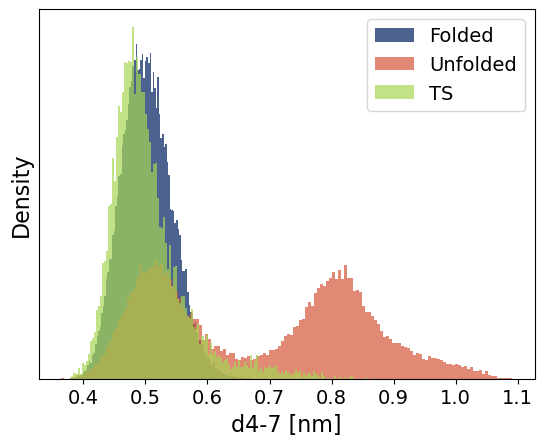}
                \caption{Normalized distribution of the distance between the $C_4^\alpha$ and $C_7^\alpha$  involved in the formation of chignolin hairpin bend sampled in the folded state, unfolded state, and the transition state (TS).}
                \label{sup_fig:chignolin_distribution_d47}
            \end{figure}

            In Fig.~\ref{sup_fig:chignolin_distribution_d47}, we report the distribution of the distance between $C_{\alpha}$ atoms 4 and 7, as a symbol of the hairpin bend involving the 4-5-6-7 residues. 
            As expected, both the folded and transition clearly peaked at $0.5$ nm, indicating the formation of the hairpin bend.
            Less obviously, the same peak, even if smaller, is also present in the unfolded state, indicating that the bending in 4-5-6-7 is a necessary but not sufficient condition for the TS. It also follows that this feature alone is not characteristic enough to identify the TSE despite being somehow intuitive.

\begin{figure}
                \centering
                \includegraphics[width=0.8\linewidth]{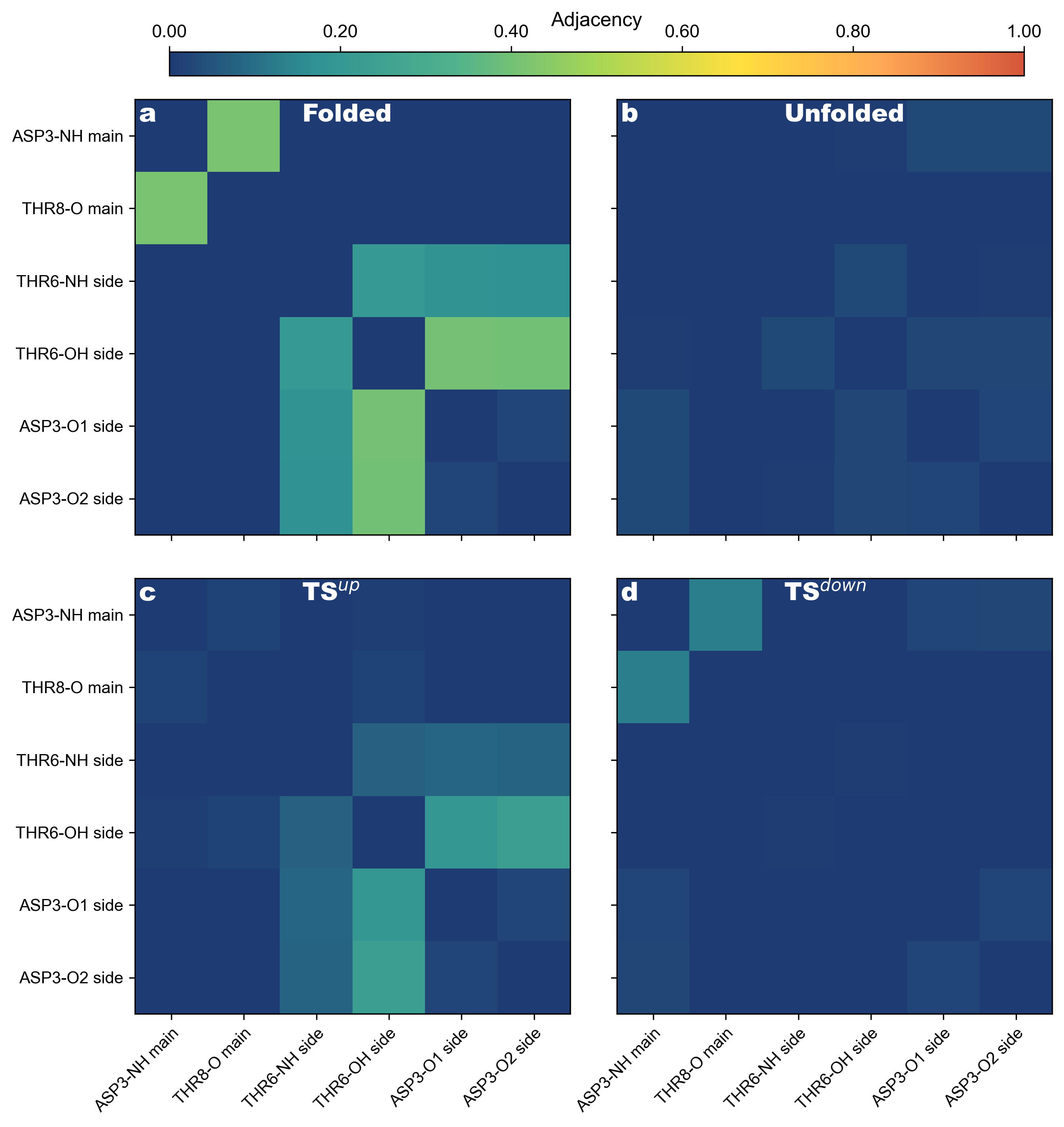}
                \caption{Average values of the adjacency matrix of the functional groups involved in the H-bond that stabilize the hairpin structure of chignolin. Values from the folded state (\textbf{a}), unfolded state (\textbf{b}), and the two TSE clusters (\textbf{c} and \textbf{d}) are reported and compared. }\label{sup_fig:chignolin_distribution_adjacentmatrix}
            \end{figure}

            In Fig.~\ref{sup_fig:chignolin_distribution_adjacentmatrix}, we report the average values of the adjacency matrix between the atoms from the functional groups involved in the monodentate H-bond between Asp3 and Thr8 and the bidentate H-bond between Asp3 and Thr6. 
            The results are reported for 4 different scenarios: folded state (\textbf{a}), unfolded state (\textbf{b}), TS$^{up}$ cluster (\textbf{c}), and TS$^{down}$ cluster (\textbf{d}).
            The results clearly show the role of the H-bonds discussed in the main text, with the formation of an H-bond network in the folded state that is completely missing on average in the unfolded state.
            In the TSE configurations, the network is partially formed, and two clusters can be identified based on which H-bonds are present.
            
\end{document}